\documentclass[10pt,twocolumn,prb,showpacs,amssymb,superscriptaddress]{revtex4-1}

\usepackage{graphicx,hyperref}
\usepackage{amsmath}
\usepackage{amsfonts}
\usepackage{epsfig} 
\usepackage{color}
\usepackage{ulem}

\begin{document}

\title{Universal crossover from ground state to excited-state quantum criticality}

\author{Byungmin Kang}
\affiliation{Department of Physics, University of California, Berkeley, Berkeley CA 94720, USA}

\author{Andrew C. Potter}
\affiliation{Department of Physics, University of California, Berkeley, Berkeley CA 94720, USA}
\affiliation{Department of Physics, The University of Texas at Austin, Austin, TX, 78712, USA}

\author{Romain Vasseur}
\affiliation{Department of Physics, University of California, Berkeley, Berkeley CA 94720, USA}
\affiliation{Materials Science Division, Lawrence Berkeley National Laboratory, Berkeley CA 94720, USA}

\date{\today}

\begin{abstract}
We study the nonequilibrium properties of a nonergodic random quantum chain in which highly excited eigenstates exhibit critical properties usually associated with quantum critical ground states. The ground state and excited states of this system belong to different universality classes, characterized by infinite-randomness quantum critical behavior. Using strong-disorder renormalization group techniques, we show that the crossover between the zero and finite energy density regimes is universal. We analytically derive a flow equation describing the unitary dynamics of this isolated system at finite energy density from which we obtain universal scaling functions along the crossover. 

\end{abstract}

\maketitle

\section{Introduction}
The concepts of scaling and universality near a critical point are central to the modern understanding of statistical mechanics and condensed matter physics, culminating in the development of the renormalization group. Whereas classical (thermal) phase transitions occur at finite temperatures and separate states with extensive (volume-law) entanglement, quantum mechanics also allows for the possibility of quantum critical points occurring at zero temperature separating distinct ground states with area-law entanglement~\cite{sachdev2011}. In both cases, universal features emerge at long distances and low temperature, so that the high-temperature regime of many-body quantum systems is often associated with non-universal, classical properties. This conventional wisdom relies on the assumption of thermal equilibrium, either due to coupling to an external heat bath, or because the system under consideration acts as its own heat bath and thermalizes on its own.

However, not all many-body quantum systems self-thermalize in isolation, and the laws of thermodynamics can break down in the presence of strong disorder because of the localization of excitations~\cite{PhysRev.109.1492} that would ordinarily move around and establish thermal equilibrium. Such {\it many-body localized} (MBL) systems~\cite{FleishmanAnderson,Gornyi,BAA,PhysRevB.75.155111,PalHuse} fail to act as their own heat bath and remain firmly out-of-equilibrium~\cite{2014arXiv1404.0686N}, thus opening the door to qualitatively new quantum critical, universal phenomena at very high-energy density, in a regime where ordinary, thermalizing systems would be effectively at infinite temperature. This can be understood intuitively as the highly excited eigenstates of MBL systems behave as zero-temperature quantum ground states, and in particular are characterized by an area-law scaling of the entanglement entropy~\cite{BauerNayak,PhysRevLett.111.127201}.

MBL systems raise the intriguing possibility of completely novel, excited-state phase transitions far from equilibrium. In particular, the dynamical transition between the MBL phase, characterized by an area-law structure of entanglement, and an ergodic regime with highly entangled eigenstates---as required by the eigenstate thermalization hypothesis (ETH)---is believed to be characterized by a very rich scaling structure that has attracted a lot of attention recently~\cite{PhysRevB.75.155111,PalHuse,MeanFieldMBLTransition,Luitz,VHA,PVPtransition}. Moreover, MBL systems can support various types of symmetry-breaking, topological, and symmetry protected topological orders~\cite{HuseMBLQuantumOrder,BauerNayak,PekkerRSRGX,PhysRevB.89.144201,BahriMBLSPT,PhysRevLett.113.107204, 2015arXiv150505147S,2015arXiv150600592P}, and the phase transitions between different MBL states represent new classes of non-equilibrium quantum critical behavior, occurring in highly excited states.  Even more dramatically, certain models have been shown to exhibit self-organized excited state quantum critical phases that are neither thermal nor MBL, and exhibit new universality classes distinct from any equilibrium phase transition~\cite{QCGPRL}.

The universal properties of such excited-state critical points (and critical phases) separating area-law entangled MBL phases can be efficiently captured by strong-disorder real-space renormalization group (RSRG)~\cite{PhysRevLett.43.1434,FisherRSRG1,FisherRSRG2} approaches. The starting point of such approaches is the RSRG procedure that has proven very useful to study zero-temperature random spin chains, in which strong couplings in the Hamiltonian are decimated before weaker ones, putting the spins involved in a strong bond in their local ground state. The effective disorder strength grows upon renormalization so that the resulting RSRG flows to infinite randomness and is said to be asymptotically exact, meaning that it is believed to yield exact results for universal quantities such as critical exponents.  This approach was recently generalized to target many-body excited states by observing that at each step, it is possible to project the strong bond onto an excited-state manifold~\cite{PekkerRSRGX}. The resulting excited-state RSRG (RSRG-X) iteratively resolves smaller and smaller energy gaps, corresponding to slow modes in the dynamics~\cite{VoskAltmanPRL13,PhysRevLett.112.217204}. This RSRG-X approach (and variants thereof) has been applied to a variety of systems recently, including random-bond Ising-type~\cite{PekkerRSRGX,PhysRevB.93.104205,1742-5468-2016-3-033101}, Heisenberg~\cite{QCGPRL,PhysRevB.92.184203}, XX~\cite{YichenJoel}, XXZ~\cite{PhysRevB.93.134207} and XYZ~\cite{PhysRevB.94.014205} spin chains. In particular, the RSRG-X procedure was argued to flow to infinite randomness for an infinite family of random spin chains at finite energy density, resulting in non-ergodic states called quantum critical glass (QCG)~\cite{QCGPRL}, where highly excited eigenstates in the middle of the many-body spectrum (corresponding to infinite effective temperature) have quantum critical properties usually associated with $T=0$ quantum critical ground states, including logarithmic (non-thermal) entanglement and algebraic mean correlations. 


Some of these nonergodic critical states, sometimes called critical or marginal MBL in the literature~\cite{PhysRevB.93.104205,PhysRevB.94.014205}, can be thought of as a critical variants of MBL, describing critical points separating different MBL phases (such as the MBL paramagnet and spin glass phases in the random transverse field Ising model for example~\cite{PekkerRSRGX}). Similarly to MBL systems, marginal MBL phases can be characterized by independent ``local'' integrals of motion (``l-bits''~\cite{PhysRevLett.111.127201,PhysRevB.90.174202}) with algebraic rather than exponential tails. These marginal MBL phases have the same universal properties in highly excited states as in their ground states. However, some strongly disordered anyon chains can host more interesting QCG {\it phases} (instead of critical points) that do not admit a description in terms of l-bits, and that have universal exponents that are distinct from their zero-temperature counterparts, with finite energy density being a relevant perturbation in the renormalization group sense~\cite{QCGPRL}.

Whereas finite-temperature properties were originally obtained numerically in the case of the transverse field Ising model using a Monte Carlo sampling of the tree of eigenstates generated by RSRG-X~\cite{PekkerRSRGX}, it was later argued that the properties of highly excited states at infinite effective temperature can be efficiently captured analytically by exact RG flow equations~\cite{QCGPRL}. However, the nature of the finite energy density crossover and the possibility of writing down analytic flow equations for eigenstates at a given energy density were left as open questions. 

In this paper, we investigate the finite energy density properties of a random one-dimensional chain of Fibonacci anyons~\cite{PhysRevLett.98.160409}, first introduced in the context of topological quantum computation~\cite{RevModPhys.80.1083}. This anyon chain is the simplest example of QCG where the infinite temperature universality class~\cite{QCGPRL} is different from that of the ground state~\cite{Bonesteel,FidkowskiPRB08,FidkowskiPRB09}. We argue here that such disordered Fibonacci anyons at arbitrary non-zero energy density (the analog of ``finite temperature")
can be described by analytic flow equations, thereby avoiding a computationally approximate numerical result via expensive Monte Carlo sampling. We justify our approach numerically and we use these exact flow equations to extract the universal features of the finite energy density crossover.  

Before going farther, it is worth clarifying what we mean by finite temperature. Just like an MBL system, we note that coupling a QCG to an external heat bath would eventually lead to delocalization~\cite{PhysRevB.90.064203}. However,  our finite-temperature flow equations can be used to describe the thermal response (say the ac thermal conductivity) of a QCG that was initially coupled to a bath and where the coupling to the bath was turned off as explained in Ref.~\onlinecite{PekkerRSRGX}. Alternatively, one can think of sampling many-body eigenstates according to a Boltzmann distribution in order to target the properties of eigenstates at a specific energy density. A natural experimental setup would indeed be to monitor the nonequilibrium dynamics of a wave packet centered on a given energy density, with a width vanishing in the thermodynamic limit. In the absence of additional information, targeting this energy density using a Boltzmann sampling of the eigenstates is the most natural choice (as the distribution maximizing the entropy). For instance, the dynamics after a global quench starting from a completely random state with no other prior information should be well described by averaging over the full many-body spectrum with a uniform distribution ($T=\infty$), corresponding for entropic reasons to the properties of typical eigenstates in the middle of the many-body spectrum. 

The remainder of this paper is organized as follows: In Sec.~\ref{SecDisorderedFibo}, we introduce a simple model of a disordered, interacting one-dimensional system, the so-called Fibonacci chain, for which the zero and infinite temperature properties are different, and we introduce a real-space renormalization group (RG) method to understand the properties of this 1D system in isolation. In Sec.~\ref{secRSRGX}, we discuss different ways to target states with a given energy density (or ``temperature''); this allows us to derive analytically RG flow equations at finite temperature or energy density. We then analyze in Sec.~\ref{flow_equation} the universal crossover to between the zero and infinite temperature limits. 


\section{Disordered Fibonacci chain at $T=0$}
\label{SecDisorderedFibo}

\subsection{Golden chain}

We begin by introducing a simple one-dimensional model, the so-called Fibonacci chain (also called the golden chain), whose eigenstates exhibit quantum critical properties, with different critical exponents and universal classes in the ground state and in highly excited states. This ``spin'' chain is constructed in analogy with the spin-$\frac{1}{2}$ Heisenberg model $H =  \sum_i J_{i} \vec{S}_i \cdot \vec{S}_{i+1} $ for which the nearest neighbor interaction gives a different energy to the singlet and triplet channels in the tensor product (fusion) $ \frac{1}{2} \otimes \frac{1}{2} = 0 \oplus 1 $, where we labeled the irreducible representations of $SU(2)$ by their spin $j=0,\frac{1}{2},1,\dots$. Whereas the $SU(2)$ group has infinitely many irreducible representations, it is useful to think of models similar to the spin-$\frac{1}{2}$ Heisenberg chain where the number of ``spins'' is truncated. The Fibonacci chain is an example of such truncation of the Heisenberg chain, and models a chain of non-Abelian anyons carrying a nontrivial topological charge $\tau$, with the fusion property 
 \begin{equation}
 \tau \otimes \tau = 1 \oplus \tau.
 \label{eqFusiontau}
\end{equation}
In analogy with the Heisenberg chain, the Hamiltonian of such a system would then take the form
\begin{equation}
H = -\sum_{i=1}^N J_iP_i^A 
\label{Hamiltonian_disordered_Fibonacci}
\end{equation}
where $J_i$ are the interaction strength which are drawn from some random distribution and $P_i^A$ is the singlet projection operator (projection to the trivial anyon) between site $i$ and $i+1$. Contrary to the fusion rule for $SU(2)$ spins $\frac{1}{2}$, 
Eq.~\ref{eqFusiontau} 
implies that the Hilbert space of $N+1$ anyons is equal to the $N$th Fibonacci number $F_N$, where $F_1=1=F_2$, and $F_{n+1}=F_n+F_{n-1}$ which scales asymptotically as $\varphi^N$ where $\varphi = \frac{1+\sqrt{5}}{2}$ is the golden ratio. This implies that each Fibonacci anyon has quantum dimension, $\varphi$, an irrational number. This implies that the Hilbert space of the Fibonacci chain cannot be described as a tensor product of local degrees of freedom. The precise definition of the Hilbert space and of the expression of the Hamiltonian~\eqref{Hamiltonian_disordered_Fibonacci} is reviewed in Appendix~\ref{Appendix_A}, in relation with the more general $SU(2)_k$ anyonic chains (the Fibonacci case corresponding to $k=3$). Beyond the potential relevance of this one-dimensional Hamiltonian in the context of topological quantum computation~\cite{RevModPhys.80.1083} (where it could be realized as a quasi-one-dimensional trench of anyonic excitations of a topologically ordered phase), we find it useful to think of anyonic chains such as~\eqref{Hamiltonian_disordered_Fibonacci} as a convenient lattice regularization of critical points that would be multi-critical (and therefore highly fine-tuned) in regular spin chains. For example, at zero temperature anyonic chains with uniform couplings are known to provide a very natural lattice regularization of the minimal models of conformal field theory~\cite{Trebst01062008}, and their ground states in the presence of strong disorder correspond~\cite{FidkowskiPRB09} to the so-called Damle-Huse infinite randomness fixed points~\cite{DamleHuse}.


\subsection{Real space renormalization group at $T=0$}
\begin{figure}[t!]
\includegraphics[width=1.0\columnwidth]{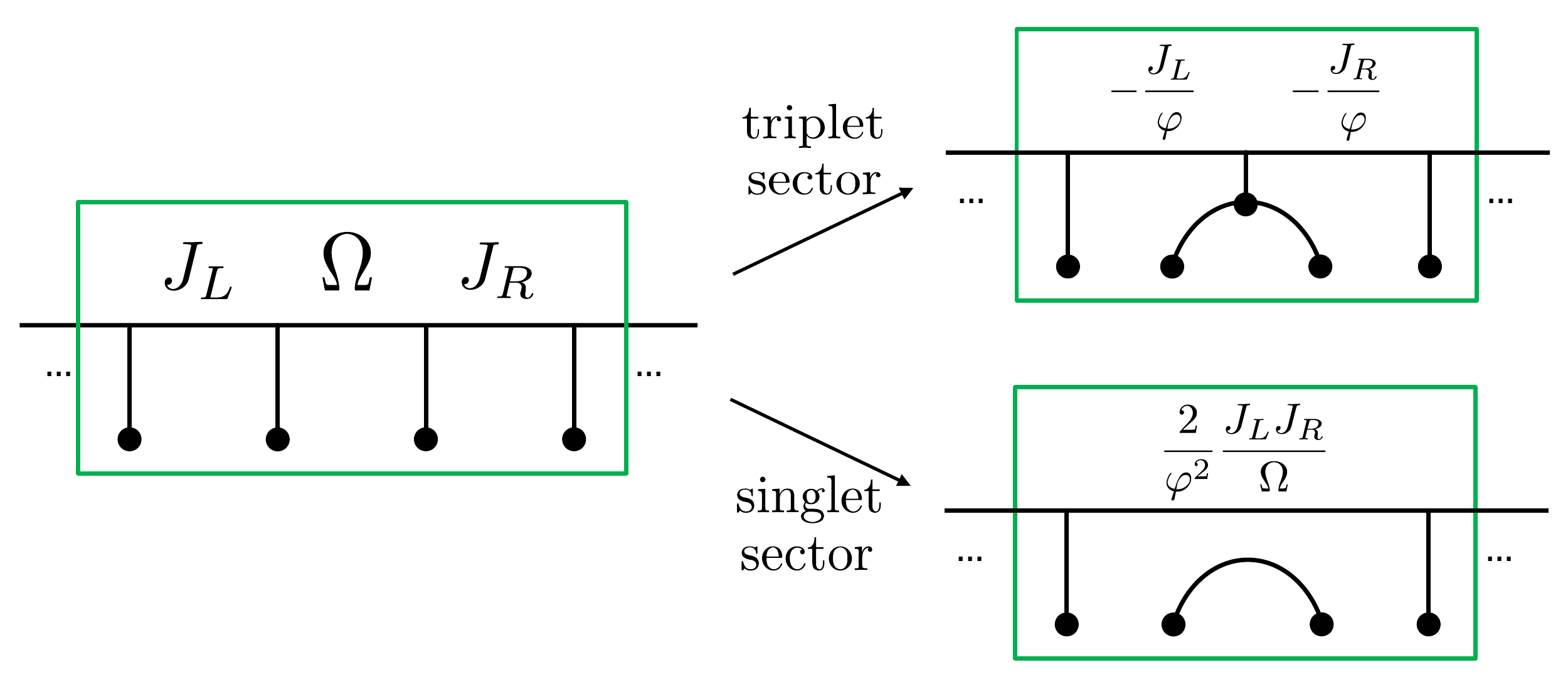}
\caption{Decimation rules for the disordered Fibonacci anyon chain. We decimate $-J_i P_i^A$ and set $J_L = J_{i-1}$ and $J_R = J_{i+1}$. Left: Strong bond in the chain. Right: The Hamiltonian after the decimation. The triplet (singlet) sector decimates 1 (2) spin(s) and renormalizes the Hamiltonian.}
\label{Fig_Fibonacci_RSRG}
\end{figure}

 Similar to ordinary disordered spin chains~\cite{PhysRevB.22.1305,PhysRevLett.43.1434,FisherRSRG1,FisherRSRG2}, the ground state ($T=0$) properties of the disordered Fibonacci chain~\eqref{Hamiltonian_disordered_Fibonacci} can be obtained~\cite{Bonesteel,FidkowskiPRB08,FidkowskiPRB09} via strong-disorder, real-space renormalization group (RSRG) methods. The key idea of this approach is to focus on the strongest bond of the chain $\Omega=\left| J_i \right|$, which at strong disorder will be typically much larger than its neighbors $\Omega \gg \left|J_{i+1}\right|,\left|J_{i-1}\right|$. We can then diagonalize this strong bond by choosing the singlet channel for an antiferromagnetic (AF) strong coupling $J_i>0$, or the triplet channel for a ferromagnetic (FM) strong coupling $J_i<0$. In the singlet channel, the two Fibonacci anyons on sites $i$ and $i+1$ form a singlet and quantum fluctuations induce an effective second-order coupling $J_{\rm eff} = \frac{2}{\varphi^2} \frac{J_L J_R}{J_{i}^2} $ between the anyons on sites $i-1$ and $i+2$. In the triplet channel, the  Fibonacci anyons on sites $i$ and $i+1$ instead form an effective new Fibonacci anyon that interacts with its neighbors via the first order couplings $- \frac{J_{L/R}}{\varphi}$. These decimation rules conserve the form of the original Hamiltonian and are summarized in Fig.~\ref{Fig_Fibonacci_RSRG}. While this RSRG is {\it a priori} approximate and accurate only at strong disorder, the effective disorder strength increases along the RG flow so that this method becomes ``asymptotically exact'' and yields exact predictions for universal properties such as critical exponents.  
 
This RSRG approach predicts that the ground state of the random Fibonacci chain is in a random-singlet phase, an infinite randomness quantum critical point characterized by algebraically decaying averaged correlation functions~\cite{FisherRSRG1,FisherRSRG2,PhysRevB.51.6411}, logarithmic scaling of entanglement~\cite{RefaelMoore}, and energy-length scaling~\cite{FisherRSRG2,PhysRevB.51.6411} 
\begin{equation}
\ln \frac{1}{E} \sim L^\psi,
\label{eqScalingtE}
\end{equation}
instead of the usual quantum-critical relation $E\sim L^{-z}$. The exponent $\psi$ in Eq.~\eqref{eqScalingtE} is given by $\psi=\frac{\Pi_s}{1+\Pi_s}$, where $\Pi_s$ is the probability that an RG step results in a second-order decimation by fusing two anyons into a singlet.
For an antiferromagnetic chain~\cite{Bonesteel}, all decimated bonds are fused to a singlet $\Pi_s=1$, so that $\psi_{\rm AF} = \frac{1}{2}$. Introducing a finite fraction of ferromagnetic bonds can be argued to be a relevant perturbation to this AF fixed point~\cite{FidkowskiPRB08}, and the system flows to a fixed point~\cite{DamleHuse} with equal ratio of F and AF bonds, so that $\Pi_s=\frac{1}{2}$ and $\psi_{\rm F/AF}=\frac{1}{3}$.

\section{Finite temperature real-space renormalization group}
\label{secRSRGX}
Whereas the discussion above focused on constructing iteratively the ground state of the Fibonacci chain, Eq.~\eqref{Hamiltonian_disordered_Fibonacci}, we are interested in this paper in the quantum critical behavior of finite energy density excited states. For this purpose, we introduce a variation of the RSRG introduced above called RSRG-X~\cite{PekkerRSRGX}, designed to target excited states. This method was first introduced as a numerical method to analyze the finite energy density properties of the disordered Ising spin chain, and it was soon after generalized to disordered anyon chains~\cite{QCGPRL} like Eq.~\eqref{Hamiltonian_disordered_Fibonacci} that show a much richer critical behavior (whereas the excited states in the Ising model have the same properties as the ground state~\cite{PekkerRSRGX,PouranvariPRB15}). Going beyond the numerical analysis of Ref.~\onlinecite{PekkerRSRGX}, it is possible to write down exact analytic flow equations to describe the behavior of infinite-temperature eigenstates~\cite{QCGPRL} ({\it i.e.} eigenstates deep in the middle of the many-body spectrum which would have $T=\infty$ in a thermal system), similarly to the ground state case ($T=0$) described above. Here, our goal is to analyze the crossover from $T=0$ and $T=\infty$ analytically by deriving flow equations to target finite temperature eigenstates. Before proceeding, we first review the RSRG-X method and the exact solution at $T=\infty$. 

\subsection{Infinite temperature quantum critical glass}

\begin{figure}[t!]
\includegraphics[width=1.0\columnwidth]{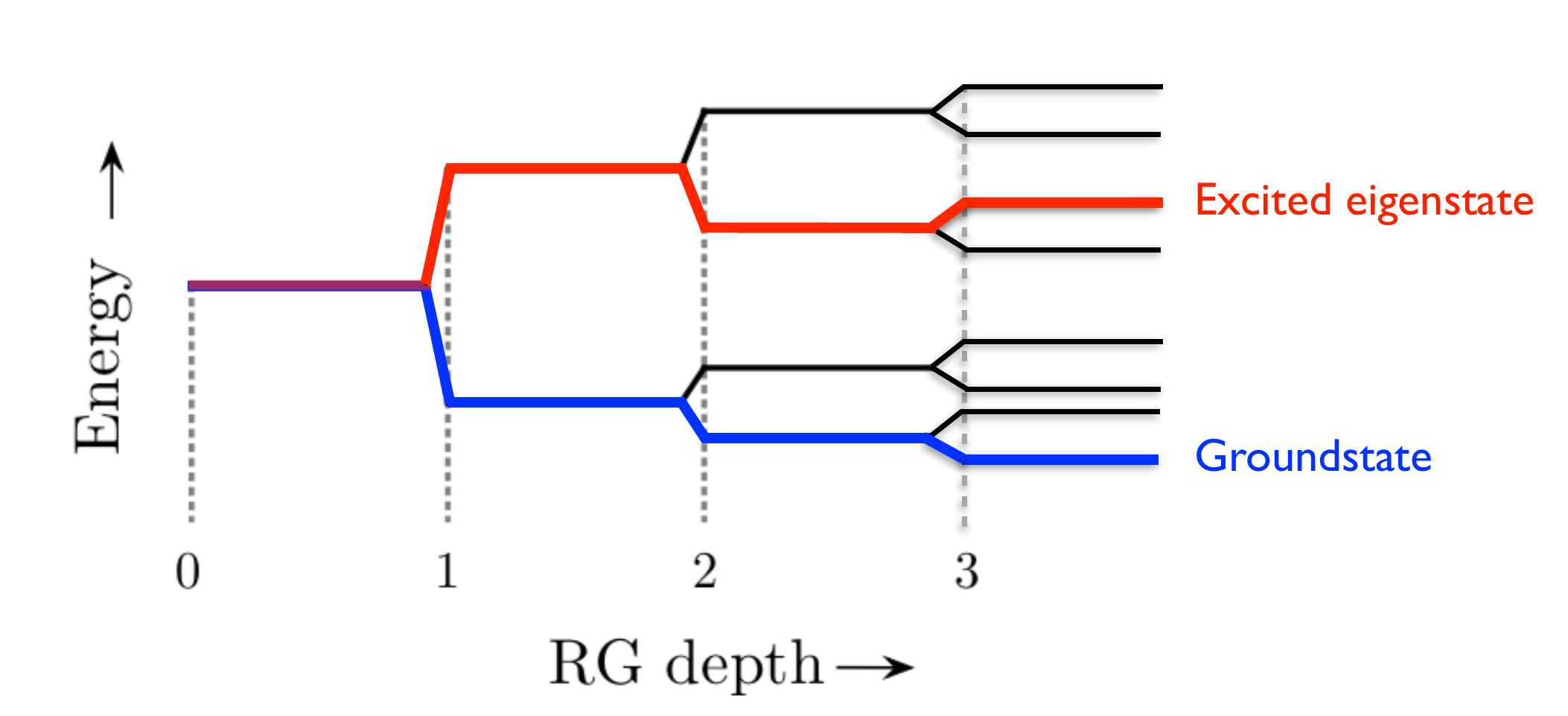}
\caption{Schematic tree of many-body eigenstates generated by RSRG-X. Every eigenstate corresponds to a branch in this tree of choices between singlet and triplet channels at each decimation. The ground state branch (blue path) is chosen by minimizing the energy at each RG step, whereas typical many-body eigenstates at $T=\infty$ (red path) can be obtained by performing a random walk on the tree.  }
\label{figRSRGXtree}
\end{figure}

The ground-state RSRG described above can be straightforwardly generalized to target many-body excited states by observing that at each step, it is possible to project the strong bond onto an excited-state manifold~\cite{PekkerRSRGX}. The resulting excited-state RSRG (RSRG-X) iteratively resolves smaller and smaller energy gaps, corresponding to slow modes in the dynamics~\cite{VoskAltmanPRL13,PhysRevLett.112.217204}, and allows one to construct, in principle, all the many-body eigenstates of the system~\cite{PekkerRSRGX,QCGPRL} (see also~\onlinecite{1742-5468-2016-3-033101,PhysRevB.93.104205,PhysRevB.93.134207}). For our example of the Fibonacci chain Eq.~\eqref{Hamiltonian_disordered_Fibonacci}, one identifies and diagonalizes the strongest bond $\Omega$ in the chain just like the ground-state version. 
However, instead of automatically choosing the fusion channel with minimal energy, one can choose to sometimes increase the total energy by choosing higher energy fusion channels.
The whole many-body spectrum can in principle be obtained in this way, and every eigenstate corresponds to a ``branch'' in the tree of choices between singlet and triplet channels at each step of the RG~(Fig.~\ref{figRSRGXtree}). For instance, the ground state (the zero-temperature sampling~\cite{PhysRevB.22.1305,PhysRevLett.43.1434,FisherRSRG1,FisherRSRG2,FidkowskiPRB08}) can be obtained by following the branch with the lowest energy, i.e., by choosing the singlet (triplet) channel if the strong bond is antiferromagnetic (ferromagnetic).

Even if RSRG-X allows one to construct in principle all the many-body eigenstates, computing physical quantities remains a very complicated task due to the exponential size of the Hilbert space. However, $T=\infty$ properties can be simply obtained by noting that for entropic reasons, typical eigenstates in the middle of the many-body spectrum can be targeted by performing a random walk on the tree of many-body eigenstates~\cite{QCGPRL}. This amounts to choosing a fusion channel (or a branch) with a probability based on the dimension of the Hilbert space of the effective chain after decimation. In the Fibonacci chain with $N+1$ anyons, this means we should weight the singlet channel with probability $F_{N-2}/F_N$ and the triplet branch with probability $F_{N-1}/F_{N}$ to ensure that each branch (or each eigenstate) is given the same probability by the RG ($T=\infty$). When $N$ is large, these probabilities converge to $1/\varphi^2$ and $1/\varphi$, respectively. 
From now on, for notational convenience we will fix $\Pi_s = 1/\varphi^2$ to be the singlet fusion probability at infinite temperature. The critical behavior of $T=\infty$ eigenstates is therefore given by a RG flow similar to $T=0$, but where
the probability to fuse to a singlet at each RG step is governed by entropy maximization rather than energy minimization. The $T=\infty$ RSRG-X still flows to an infinite-randomness fixed point characterized by the scaling~\eqref{eqScalingtE}, but with a new infinite-temperature tunneling exponent: 
\begin{equation}
\psi_\infty = \frac{1}{2+\varphi},
\end{equation}
different from the $T=0$ exponents $\psi_{\rm AF}=\frac{1}{2}$ and $\psi_{\rm F/AF}=\frac{1}{3}$. Dimerization is irrelevant at the infinite-temperature fixed point~\cite{QCGPRL} (this is related to the impossibility of many-body localizing Fibonacci anyons~\cite{PhysRevB.94.224206}), and the resulting $T=\infty$ critical phase was dubbed quantum critical glass (QCG) in Ref.~\onlinecite{QCGPRL}, and is characterized by a non-ergodic logarithmic scaling of eigenstate entanglement $S\sim \ln L$ violating ETH, and by algebraic decaying averaged correlation functions. Although there is no formal proof for the stability of this non-ergodic critical phase against thermalization as in the MBL case~\cite{PhysRevLett.117.027201}, several arguments~\cite{VoskAltmanPRL13,PVPtransition} suggest it is stable at strong enough disorder, while for weak disorder, the proliferation of many-body resonances ignored by RSRG-X naturally leads to thermalization~\cite{PhysRevB.94.235122}. Regardless of this question of stability of QCG in the thermodynamic limit, we remark that RSRG-X can be made arbitrarily accurate for any finite size system by increasing the disorder strength.   

We emphasize that this QCG nonergodic phase is very different from ``marginal MBL'' critical points that arise in the random transverse field Ising chain for example, for which the ground-state critical behavior simply extends to finite energy density. In the random Fibonacci chain, the ground state  and excited states of the system belong to completely different universality classes, characterized by distinct RG fixed points. Moreover, random Fibonacci anyons are generically critical and cannot be many-body localized~\cite{PhysRevB.94.224206}---even by strongly dimerizing the couplings~\cite{QCGPRL}---so that the QCG state should be considered as a (nonequilibrium) phase, rather than a fine-tuned critical point.   

\begin{figure}[t!]
\includegraphics[width=0.8\columnwidth]{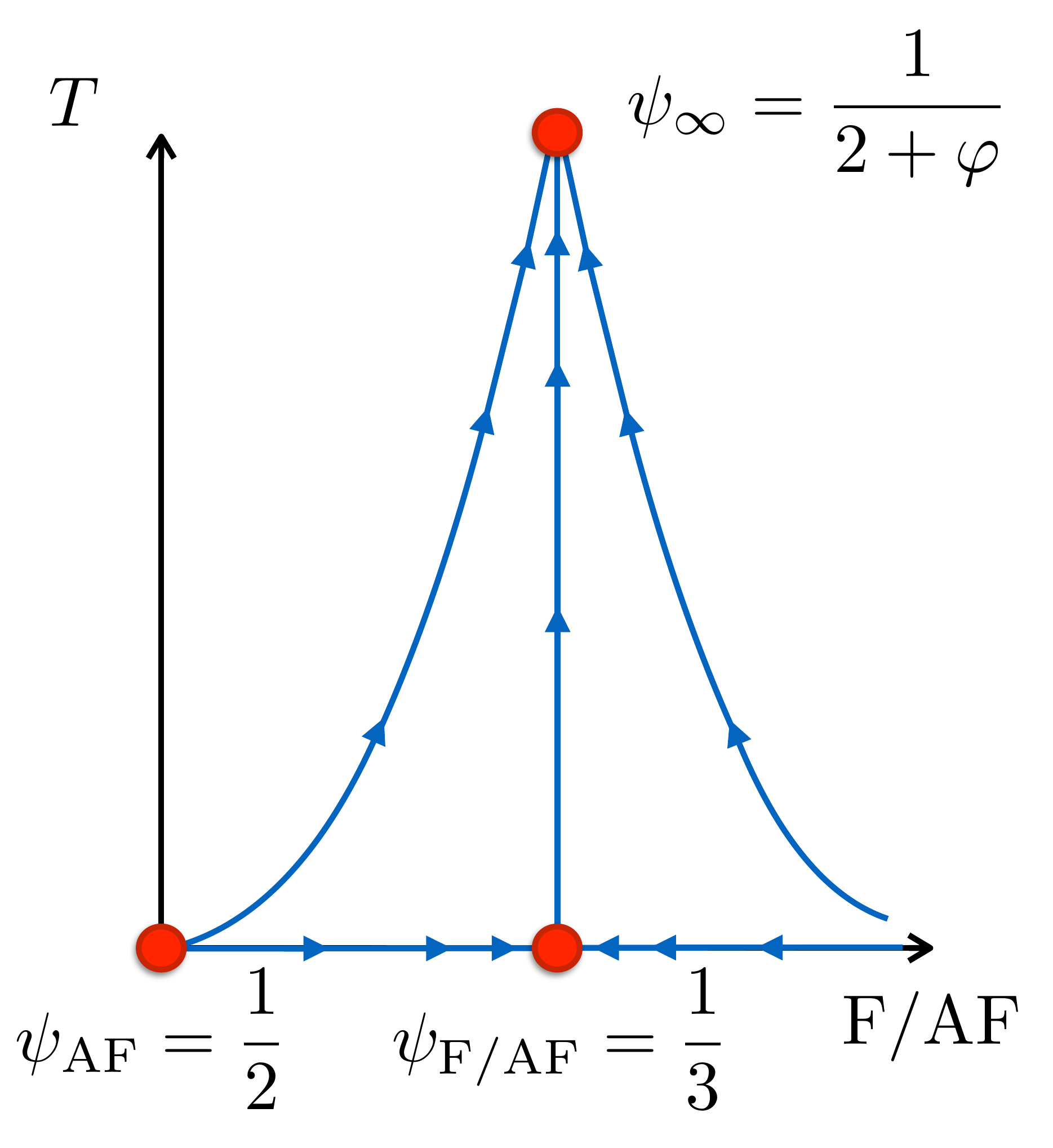}
\caption{Excited-state RG flow for the random Fibonacci chain. At zero temperature, the system is described by two infinite-randomness random singlet fixed points depending on the ratio of ferromagnetic bonds in the chain, with different $\psi$ exponents characterizing the scaling between length and energy $\Gamma= - \ln E \sim L^\psi$. Considering the system at finite energy density or temperature is a relevant perturbation in the renormalization group sense, and the system flows to a stable quantum critical glass (QCG) fixed point. In this paper, we analyze the universal finite-temperature crossover between the $T=0$ and $T=\infty$ fixed points.    }
\label{Fig_RGFlow}
\end{figure}

As we will see in detail in the following, finite temperature or energy density is a relevant perturbation to the $T=0$ fixed points, so that the system flows to the $T=\infty$ QCG fixed point for all temperatures $T>0$. The corresponding schematic RG flows are sketched in Fig.~\ref{Fig_RGFlow} and the goal of this paper is to study the (universal) finite-temperature crossover between the $T=0$ and $T=\infty$ behaviors. We now turn to how finite-temperature properties can be efficiently obtained using RSRG-X.

\subsection{Local node sampling vs.  Monte Carlo sampling}
\label{subsecLocalnode}
As we argued in the introduction, a physically motivated way to interpolate between the $T=0$ and $T=\infty$ quantum critical behaviors is to weight finite energy density eigenstates by a Boltzmann distribution. We emphasize however that we are dealing with an isolated non-ergodic system, so that the effective temperature $T$ is used merely as a way to target the properties of finite energy density eigenstates. The expectation is that typical eigenstates with similar energy density, $\epsilon$, have the same bond distribution, and since the Boltzmann distribution produces uncertainty in energy density that scales like $\Delta\epsilon \sim 1/\sqrt{L}$ where $L$ is the system size, that this sampling method enables us to reproduce the dynamical scaling properties of a more generic quench from an arbitrary non-eigenstate with typical energy density $\epsilon$.
Within RSRG-X, this means that we want to sample eigenstates with energy $E$ according to a Boltzmann distribution ${\rm e}^{-\beta E}/Z$. 

In Ref.~\onlinecite{PekkerRSRGX}, it was demonstrated that this could be done numerically using a Monte Carlo sampling using the Metropolis algorithm~\cite{metropolis1953equation}. This Monte Carlo sampling of the RSRG-X tree of eigenstates can be summarized as follows: let us start with a single eigenstate (corresponding to a given branch in the tree), which we call a sample eigenstate, with energy $E_s$. We make a trial move by picking a random eigenstate, a trial eigenstate, with energy $E_t$. We accept the move if $e^{-\beta E_t}/e^{-\beta E_s}$ is larger than a random number in an interval $[0,1]$. If the trial state is accepted, then the trial state becomes a sample eigenstate and if not, then the sample eigenstate remains the same. At the end of each attempt, we compute some desired quantities, e.g., the entanglement entropy or some correlation function, and because of the detailed balance condition we are guaranteed to sample eigenstates according to the Boltzmann distribution at $\beta$. Even though this Monte Carlo sampling is a powerful tool that allows one to partly overcome the exponential number of eigenstates, generating a random eigenstate requires some computation time and limits the system size to ${\it O}(100)$ sites. This number might be too small to capture the fixed point coupling distribution which would require at least $10^4$ spins. Another disadvantage of the Monte Carlo sampling is that it gets harder upon approaching zero temperature because, since each trial eigenstate is chosen randomly, it is exponentially hard to sample exactly the ground state at zero temperature as the system size becomes larger. 

Because of these complications, we seek an analytically tractable alternative to Monte Carlo sampling. To this end, we note that a sampling method that leads to analytically tractable flow equations is to independently weight the fusion choice locally at each node of the RSRG-X tree~(Fig.~\ref{figRSRGXtree}), rather than globally based on the entire energy of the branch. Namely, consider the following local-node weighting scheme. If the strongest bond is antiferromagnetic, $J_i>0$, then we choose the singlet branch with probability $p_{A}^1 = \frac{\Pi_s e^{\beta \Omega}}{(1-\Pi_s) + \Pi_s e^{\beta \Omega}}$ and choose the triplet branch with probability $p_{A}^\tau = \frac{1-\Pi_s}{(1-\Pi_s) + \Pi_s e^{\beta \Omega}}$. Otherwise, if the strongest bond is ferromagnetic, $J_i<0$, then we choose the singlet branch with probability $p_{F}^1 = \frac{\Pi_s e^{-\beta \Omega}}{(1-\Pi_s) + \Pi_s e^{-\beta \Omega}}$ and choose the triplet branch with probability $p_{F}^\tau = \frac{1-\Pi_s}{(1-\Pi_s) + \Pi_s e^{-\beta \Omega}}$. By sending $\beta$ either to infinity or zero, this ``local-node sampling'' reduces to the zero or infinite temperature cases, and at first glance, one might hope that it correctly interpolates between these limits giving the desired Boltzmann sampling of many-body eigenstates. However, this local-node sampling turns out to actually deviate from the desired Boltzmann distribution on eigenstates, as we will argue shortly for a simple four-site system. Nonetheless, we will demonstrate that local-node sampling actually provides a very good approximation at strong disorder, and hence will ultimately be successful for computing universal scaling properties due to the flow to strong disorder.


\begin{figure}[t!]
\includegraphics[width=0.8\columnwidth]{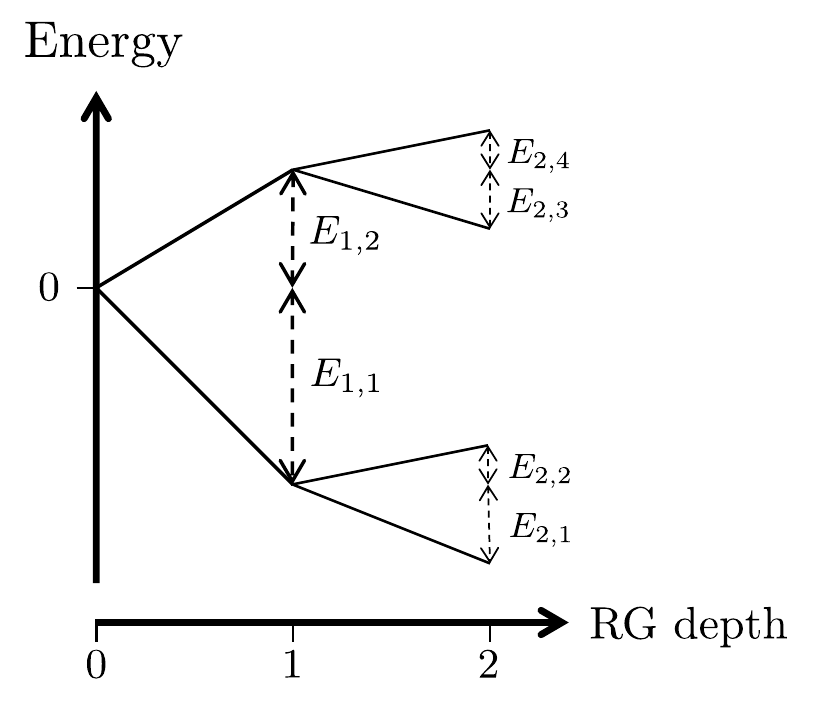}
\caption{Simplified RSRG-X tree with four eigenstates (branches) illustrating the difference between Monte Carlo and local-node finite-temperature samplings (see text).}
\label{Fig_RG_simple_example}
\end{figure}

To see that the local-node weighting scheme deviates from the desired global-branch Boltzmann weighting for finite temperatures $0<T<\infty$, let us consider a
simplified 4-state model where 4 eigenstates correspond to 4 branches at the end of the RG-tree. This model is different from the Fibonacci anyon chain and would arise for instance for an Ising chain, but it captures the essence of the local node sampling in the most simplified fashion. Our model system is described in Fig.~\ref{Fig_RG_simple_example}. The energies are given by $E_{1,1} + E_{2,1}$, $E_{1,1}+E_{2,2}$, $E_{1,2}+E_{2,3}$, and $E_{1,2}+E_{2,4}$ with the associated Boltzmann probabilities $\frac{1}{Z} e^{-\beta (E_{1,1} + E_{2,1})}$, $\frac{1}{Z} e^{-\beta (E_{1,1} + E_{2,2})}$, $\frac{1}{Z} e^{-\beta (E_{1,2} + E_{2,3})}$, and $\frac{1}{Z} e^{-\beta (E_{1,2} + E_{2,4})}$, where $Z = e^{-\beta (E_{1,1} + E_{2,1})} + e^{-\beta (E_{1,1} + E_{2,2})} + e^{-\beta (E_{1,2} + E_{2,3})} + e^{-\beta (E_{1,2} + E_{2,4})} = e^{-\beta E_{1,1}} (e^{-\beta E_{2,1}} + e^{-\beta E_{2,2}}) + e^{-\beta E_{1,2}} (e^{-\beta E_{2,3}} + e^{-\beta E_{2,4}})$ is the partition function. The probability of choosing the ground state in the local node sampling is given by the product of choosing a lower branch in RG step 1, $p_1$, and the probability of choosing a lower branch in RG step 2, $p_2$. Note that the dimension of the Hilbert space of a lower and an upper branch are the same at each RG step; we get $p_1 = \frac{e^{-\beta E_{1,1}}}{e^{-\beta E_{1,1}} + e^{-\beta E_{1,2}}}$ and $p_2 = \frac{e^{-\beta E_{2,1}}}{e^{-\beta E_{2,1}} + e^{-\beta E_{2,2}}}$. Hence the local node probability associated with the ground state is $p_1 p_2 = \frac{1}{Z'} e^{-\beta(E_{1,1} + E_{2,1})}$, where $Z' = e^{-\beta E_{1,1}} (e^{-\beta E_{2,1}} + e^{-\beta E_{2,2}}) + e^{-\beta E_{1,2}} (e^{-\beta E_{2,1}} + e^{-\beta E_{2,2}})$ which is \textit{different} from $Z$. All the other eigenstates suffer from the same problem, resulting from the fact that $E_{2,1},E_{2,2} \neq E_{2,3},E_{2,4} $, meaning that the energies involved in the second decimation are influenced by the outcome of the first decimation.

\begin{figure}[t!]
\includegraphics[width=1.0\columnwidth]{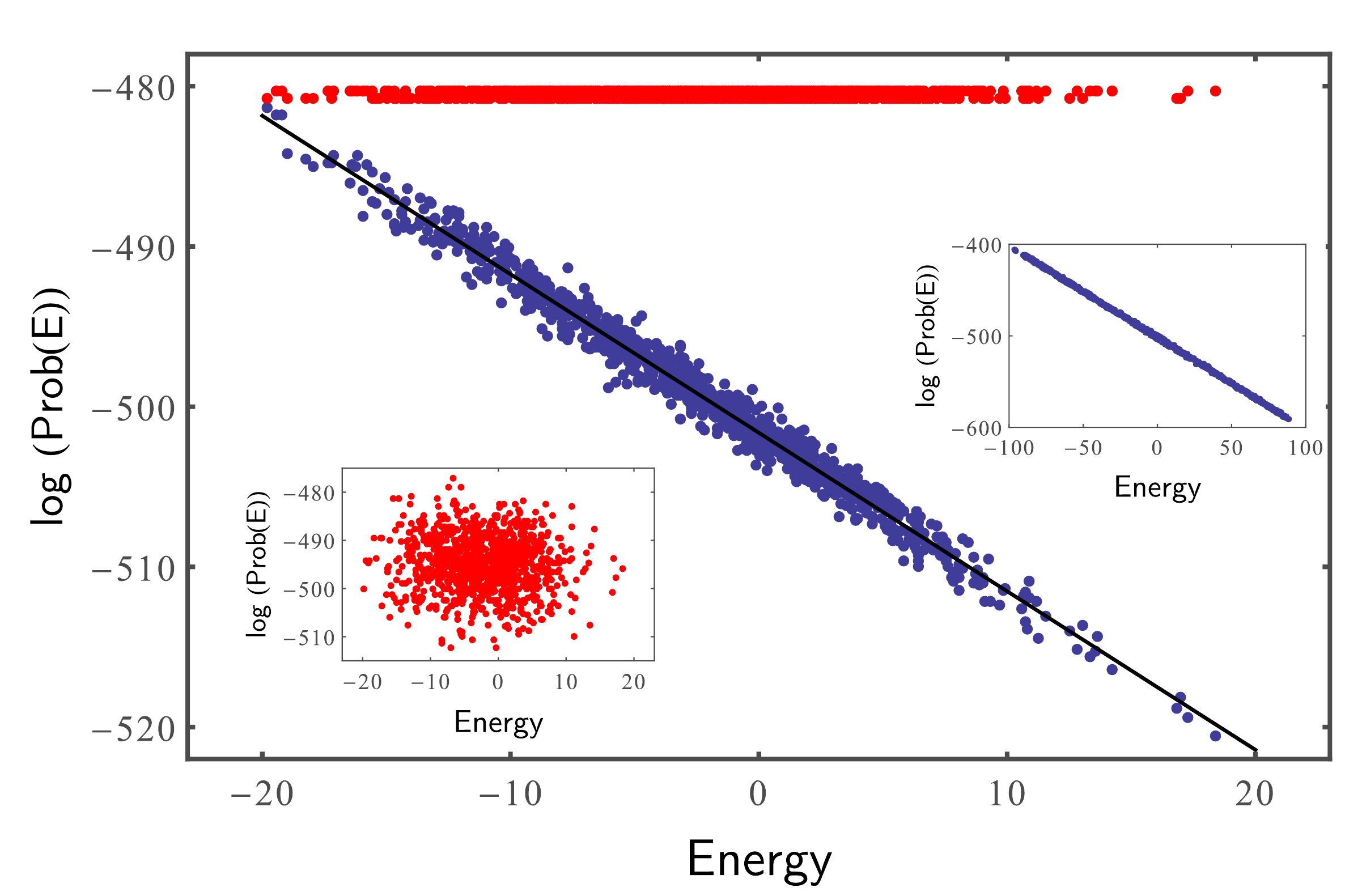}
\caption{Local-node probability versus energy of $10^3$ random eigenstates from a disordered Fibonacci chain with $10^3$ spins with the initial disorder strength $W=2$. Probabilities are calculated from the local-node probabilities with $\beta$ equals $1$ (blue) and $0$ (red). The splitting in the $\beta=0$ case is due to the finite size effect (i.e., goes away when taking number of spins to infinity), which is completely removed when using the exact finite-size single fusion probability, $F_{N-2}/F_N$, instead of $\Pi_s$. The fitted linear curve has a slope $\approx 0.99$. Left inset: Plot of local-node sampling probabilities at $\beta=0$ but with a wrong $\Pi_s$ ($0.1$ larger than the correct one) showing a nonphysical probability distribution. Right Inset: Plot of local-node sampling probabilities at $\beta=1$ sampled over the full spectrum.}
\label{Fig_log_E_vs_E}
\end{figure}

\subsection{Validity of the local node sampling and history dependence}

Despite this discrepancy, this local node sampling approximation can be justified analytically by analyzing the history dependence of the RG flow, that is, by quantifying how much  a choice of fusion channel at energy scale $\Gamma_0$ affects the rest of the RG flow for $\Gamma >\Gamma_0$. As we have seen above, the difference between the local node and the true Boltzmann samplings comes from this history dependence: if the choice of fusion channel in the first step affects the energy at a later step (in our example, $E_{2,1},E_{2,2} \neq E_{2,3},E_{2,4} $), then the two samplings will be different. In practice though, a given decimation will not affect most of the subsequent RG steps that will most likely involve bonds far away from it, but it will modify the energy when the renormalized bonds resulting from the original decimation are decimated again. 

When one decimates a strong bond with strength $ \Omega_0 =e^{-\Gamma_0}$, there are two options: either a singlet is formed with some probability leading to a strongly weakened renormalized coupling $\sim J_L J_R/\Omega_0$ between the neighboring anyons, or a new effective anyon $\tau$ is created with two effective couplings to its neighbors given by first-order perturbation theory. If we choose the singlet channel, the renormalized second-order coupling is typically much weaker than most bonds in the chain, so that it will be decimated much later in the flow. Generalizing the ideas of Ref.~\onlinecite{RefaelMoore}, we find that it will typically be decimated at scale $\Gamma_{s} = \Gamma_0 e^{\Pi_s (2 + \Pi_s)}$ where for simplicity, we assumed that the singlet fusion probability $\Pi_s$ is constant over this range of energy scales. This will lead to a contribution of order $\Omega_{s} \sim \Omega_0^{e^{\Pi_s (2 + \Pi_s)}}$ to the energy. If on the other hand we pick the $\tau$ channel at scale $\Gamma_0$, we find that the resulting first order couplings will typically be decimated at scale $\Gamma_{\rm \tau} = \Gamma_0 e^{\Pi_s (1+ \Pi_s)}$ leading to a term in the many-body energy of order $\Omega_{\tau} \sim \Omega_0^{e^{\Pi_s (1 + \Pi_s)}}$. The crucial point is that these decimations will occur much later in the RG, and give contributions to the energy $\Omega_{1}$ or $\Omega_{\tau}$ that are negligible compared to $\Omega_0$ in the strong-disorder limit $\Omega_0 \to 0$ ($\Gamma_0 \to \infty$). This implies that in the strong-disorder limit, a given decimation will essentially not affect the subsequent flow so that the local-node sampling will become an increasingly good approximation of the Boltzmann distribution. \\

\subsection{Numerics}

\begin{figure}[ht!]
\includegraphics[width=0.8\columnwidth]{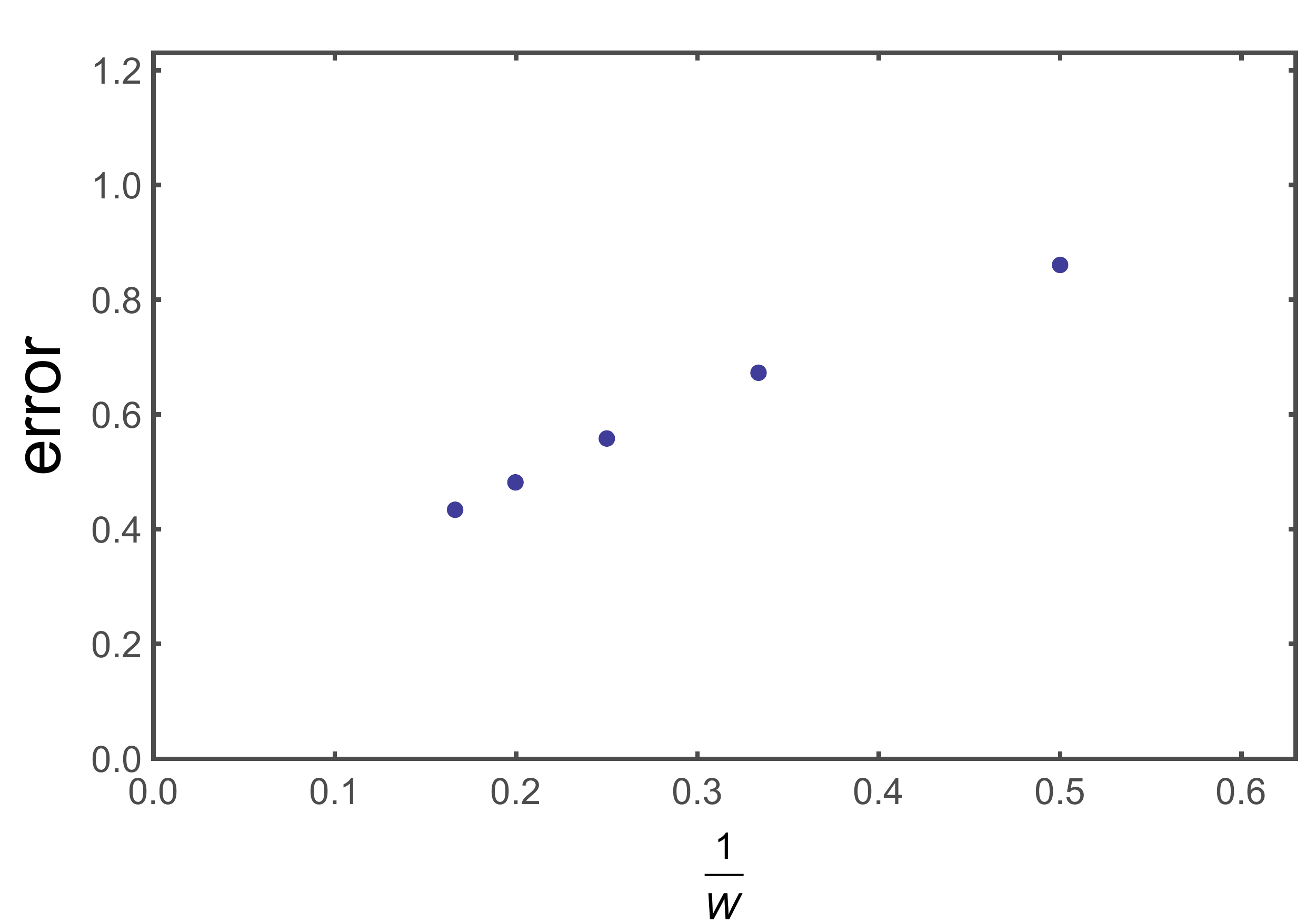}
\caption{Plot of the ``error'' in effective partition function versus the inverse disorder strength. This ``error'' is computed as the averaged value of standard deviations of effective partition functions associated with random eigenstates. We sample over $10^3$ eigenstates in each disorder realization and $500$ disorder realizations for each $W$ with $10^3$ spins. Error bars are always smaller than the size of the symbols.}
\label{Fig_W_plots}
\end{figure}

In agreement with this analytical argument, we will show numerically in the following that the local-node sampling using $(p_{A}^1, p_{A}^\tau, p_F^1, p_F^\tau)$ is a good approximation of the Boltzmann sampling with an error that decreases upon increasing the initial disorder strength. This implies that the local-node sampling using $(p_{A}^1, p_{A}^\tau, p_F^1, p_F^\tau)$ captures the correct universal long-time dynamics at finite $\beta$ since the disorder strength increases without bound along the RG flow. 


To see how the local-node sampling works in practice, we plot the probability given to eigenstates using the local node sampling as a function of energy for $10^3$ eigenstates drawn randomly in a disordered Fibonacci chain with $10^3$ anyons (Fig.~\ref{Fig_log_E_vs_E}). The initial couplings are drawn from $P(\log J^{-1}) = \frac{1}{W} e^{-\frac{\log J^{-1} }{W}}$, where $W$ is the initial disorder strength. We used RSRG-X at infinite temperature to sample eigenstates randomly and set $W=2$. The probability is then computed by a local node probability $(p_A^1,p_A^\tau,p_F^1,p_F^\tau)$ with $\beta=1$ and $0$. We find that the local node sampling gives a distribution very close to the Boltzmann result, with an effective fitted temperature $\beta \approx 0.99$ which agrees with the correct value. We remark that the data at infinite temperature ($\beta=0$) shows a small splitting instead of a perfect horizontal line because we used the expression $\Pi_s = 1/\varphi^2$ to weight the different fusion channels, which is exact in the thermodynamic limit but which differs from the exact expression $F_{N-2}/F_N$ when the number $N$ of remaining anyons becomes small near the end of the renormalization process.


To quantify the difference between Boltzmann and local-node sampling, recall that the local-node sampling associates the probability $p_i = e^{-\beta E_i}/Z_\textrm{eff}^{(i)}$ with an eigenstate $i$, where $E_i$ is the energy and $Z_\textrm{eff}^{(i)}$ is the guessed (effective) partition function of $i$. Ideally, the effective partition function should be independent of the eigenstate, and should coincide with the actual  partition function at temperature $\beta^{-1}$. To see how the accuracy of the local-node sampling changes upon increasing the initial disorder strength, we compute the averaged standard deviations of the guessed (or effective) partition function as the error as we vary the initial disorder strength $W$. We averaged over $500$ disorder realizations with $10^3$ spins and we sample $10^3$ eigenstates randomly to calculate the standard deviation of the effective partition function. It is clear from Fig.~\ref{Fig_W_plots} that the error decreases as we increase the disorder strength, which implies that the local-node sampling becomes a better approximation upon increasing the disorder strength. Since the disorder strength increases without bound along the RG flow, we expect that the local-node sampling captures the correct long-time, universal behavior of the system. We note that the nonzero value of the extrapolated error at $W=\infty$ is a finite-size artifact, and appears to vanish in the infinite-system limit. 

We note in passing that it might be possible to interpolate between the exact -- but numerically costly -- Boltzmann  (Monte Carlo) sampling  and our asymptotically exact -- approximate but analytically tractable -- local-node sampling. To be more specific, one could imagine choosing trial states of the Monte Carlo procedure by following the local-node sampling at desired temperature $\beta^{-1}$, while accepting or rejecting the trial states by using both the probability given by the exact Boltzmann factor and the probability associated with the local-node sampling. This kind of ``assisted'' Monte Carlo sampling method has appeared in the literature recently~\cite{PhysRevB.95.035105, PhysRevB.95.041101} and was been shown to be much more efficient than conventional Monte Carlo samplings. It will be interesting to investigate such improved Monte Carlo samplings in the future.

\section{Universal Finite-Temperature crossover}
\label{flow_equation}
\subsection{Finite-Temperature Flow Equations}
As we have seen in the previous section, the simple local-node Boltzmann weighting of the RG fusion branches $(p_A^1,p_A^\tau,p_F^1,p_F^\tau)$ gives arbitrary good approximation to the Boltzmann distribution at strong disorder, and can therefore be expected to accurately capture the universal finite-temperature crossover physics. Using this local node probability, we can derive analytic finite-temperature RG flow equations for the disordered Fibonacci chain in a way that is very similar to the $T=0$ and $T=\infty$ cases. These flow equations encode how the coupling distribution evolves along the RG flow. It is convenient to define the energy scale at a given RG step by $\Omega = \max_i |J_i|$, the RG scale $\Gamma = \log(\frac{1}{\Omega})$, and logarithmic couplings $\zeta_i = \log \frac{\Omega}{|J_i|}$. (We normalize the initial energy scale to be $\Omega_0 = 1$.) The full flow equations are derived in Appendix~\ref{flow_eqn_derivation}, and they read
\begin{widetext}
\begin{eqnarray}
\frac{\partial \rho_F (\zeta, \Gamma)}{\partial \Gamma} = \frac{\partial \rho_F (\zeta, \Gamma)}{\partial \zeta} &+& \rho_F(0,\Gamma) \Big[ 2 p_F^1 \rho_F \star \rho_A + 2(1-p_F^1) \rho_A (\zeta,\Gamma) - (1-p_F^1) \rho_F(\zeta,\Gamma) \Big] \nonumber \\
&+& \rho_A(0,\Gamma) \Big[ 2 p_A^1 \rho_F \star \rho_A + 2(1-p_A^1) \rho_A (\zeta,\Gamma) - (1-p_A^1) \rho_F(\zeta,\Gamma) \Big], \nonumber \\
\frac{\partial \rho_A (\zeta, \Gamma)}{\partial \Gamma} = \frac{\partial \rho_A (\zeta, \Gamma)}{\partial \zeta} &+& \rho_F(0,\Gamma) \Big[ p_F^1 \big( \rho_F \star \rho_F + \rho_A \star \rho_A \big) + 2(1-p_F^1) \rho_F (\zeta,\Gamma) - (1-p_F^1) \rho_A(\zeta,\Gamma) \Big] \nonumber \\
&+& \rho_A(0,\Gamma) \Big[ p_A^1 \big(\rho_F \star \rho_F + \rho_A \star \rho_A \big) + 2(1-p_A^1) \rho_F (\zeta,\Gamma) - (1-p_A^1) \rho_A(\zeta,\Gamma) \Big],
\label{Finite_T_flow_eqn}
\end{eqnarray}
\end{widetext}
where $\rho_A$ and $\rho_F$ are the probability distribution of the coupling strength of antiferromagnetic and ferromagnetic bonds, $p_{A}^1 = p_A^1 (\beta;\Gamma) = \frac{\Pi_s e^{\beta \Omega}}{(1-\Pi_s) + \Pi_s e^{\beta \Omega}}$ and $p_{F}^1 = p_F^1 (\beta;\Gamma) = \frac{\Pi_s e^{-\beta \Omega}}{(1-\Pi_s) + \Pi_s e^{-\beta \Omega}}$ are singlet fusion probabilities of an antiferromagentic/ferromagnetic bond at an energy-scale $\Omega = e^{-\Gamma}$, and $\rho_1 \star \rho_2 = \int_0^\zeta d\zeta' \rho_1(\zeta-\zeta',\Gamma) \rho_2(\zeta',\Gamma)$ is the convolution between the two distributions $\rho_1$ and $\rho_2$. $\rho_A$ and $\rho_F$ are normalized in such a way that $p_A = \int d \zeta \rho_A(\zeta,\Gamma)$ and $p_F = \int d \zeta \rho_F(\zeta,\Gamma)$ are the ratio of the total antiferromagnetic and ferromagnetic bonds at the RG scale $\Gamma$. Equation~(\ref{Finite_T_flow_eqn}) reduce to the zero or the infinite temperature flow equations~\cite{FidkowskiPRB08,QCGPRL} by sending $\beta$ to infinity or zero. One powerful feature of these flow equations is that they describe the flow of couplings associated with a \textit{single} typical eigenstate under the RG flow at the desired energy density (or temperature). This again shows a great advantage over the Monte Carlo sampling which requires reasonably many samplings in order to obtain finite-temperature properties. 

Like the flow equations at zero and infinite temperature, Eqs.~(\ref{Finite_T_flow_eqn}) admit a remarkably simple fixed-point (or scale-invariant) solution. Using a  stability analysis, one can show that the fixed-point distribution has equal proportion of ferromagnetic and antiferromagnetic bonds, so that $\rho_A(\zeta,\Gamma) = \rho_F(\zeta, \Gamma)$. Let us therefore define $\rho(\zeta,\Gamma) = 2 \rho_A(\zeta,\Gamma) = 2 \rho_F(\zeta,\Gamma)$, so that the flow equations (\ref{Finite_T_flow_eqn}) reduce to
\begin{equation}
\frac{\partial \rho (\zeta, \Gamma)}{\partial \Gamma} = \frac{\partial \rho (\zeta, \Gamma)}{\partial \zeta} + \rho(0,\Gamma) \Big[ \Pi_s(\beta) \rho \star \rho + (1-\Pi_s(\beta)) \rho \Big],
\end{equation}
where $\Pi_s(\beta) = \Pi_s(\beta;\Gamma) = \frac{1}{2} (p_A^1 + p_F^1)$ is the averaged singlet fusion probability. The fixed-point solution thus reads $\rho(\zeta,\Gamma) = \frac{1}{W+\Xi_s(\beta;\Gamma)} e^{-\frac{\zeta}{W+\Xi_s(\beta;\Gamma)}}$, where $\Xi_s = \int_0^\Gamma d\Gamma' \Pi_s(\beta;\Gamma')$ is the integral of $\Pi_s(\beta;\Gamma)$. Note that the Ising ($k=2)$ case has constant $\Pi_s(\beta) = 1$, i.e., the flow equation is the same as the ground state and at infinite temperature and there is \textit{no} crossover (so that there is no difference between the ground state and excited states in this case; see also~\onlinecite{YichenJoel,PouranvariPRB15}).

\begin{figure}[t!]
\includegraphics[width=1.0\columnwidth]{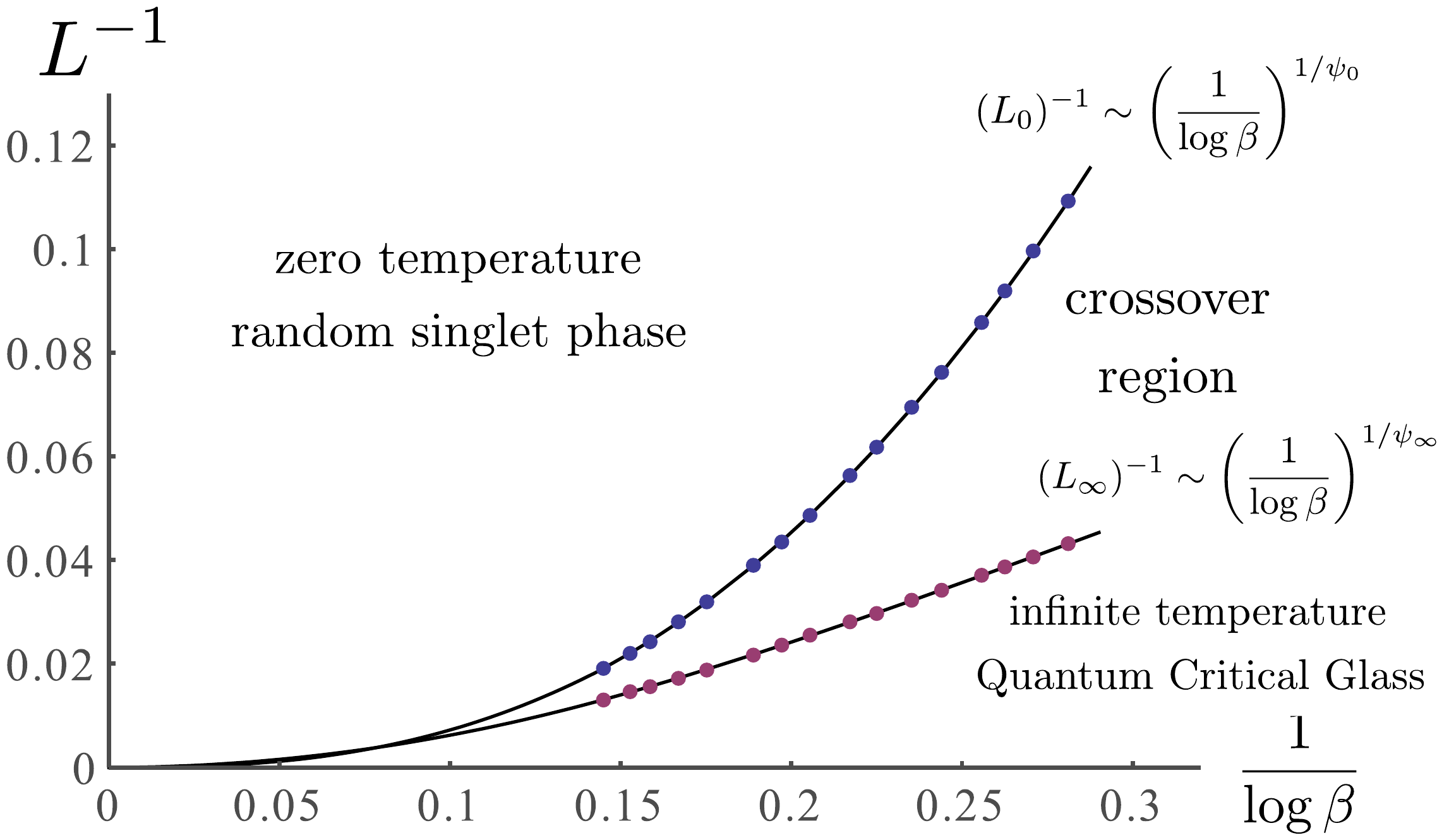}
\caption{Crossover length scales $L$ versus $\frac{1}{\log \beta}$. At small but finite temperature, the short-length scale physics is dominated by the ground-state fixed-point phase. After passing through the crossover region, the physics is eventually dominated by the infinite-temperature infinite-randomness fixed point. The crossover boundaries are determined by computing at which length scale $\Pi_s(\beta;\Gamma)$ approaches or departs from its UV (ultraviolet)/IR (infrared) value within one percent, using Eq.~(\ref{Eqn_L_Gamma}) to convert $\Gamma$ into a distance $L$.}
\label{Fig_crossover}
\end{figure}

\subsection{Universal finite-temperature crossover}

The averaged singlet fusion probability $\Pi_s(\beta;\Gamma)$ determines how fast the effective disorder strength changes as we increase the RG parameter $\Gamma$. Upon increasing $\Gamma$, $\Pi_s(\beta;\Gamma)$ exhibits a crossover from the zero-temperature value $\Pi^0_s=\frac{1}{2}$ to the infinite-temperature value $\Pi^\infty_s=\Pi_s=\frac{1}{1+\tau}$ around the energy scale $\Omega \sim 1/\beta$. We expect physical quantities such as correlation functions or entanglement entropy to show a crossover interpolating between the $T=0$ and $T=\infty$ critical properties as a function of, say, distance $L$. In order to determine how the scaling between energy and distance evolves along the crossover, we need to determine how the scaling between energy and length evolves along the flow. This can be computed by noting that as we increase the RG scale from $\Gamma$ to $\Gamma+d\Gamma$, the number of decimated spins is given by $dN = -N(\Gamma) \rho(0,\Gamma) \big(1+\Pi_s (\beta;\Gamma) \big) d\Gamma$. Using the fixed-point solution, the typical distance between spins at scale $\Gamma$ is equal to:
\begin{equation}
L(\Gamma) = \frac{N_0}{N(\Gamma)} = \exp \bigg( \int_0^\Gamma d\Gamma' \frac{1+\Pi_s (\beta;\Gamma')}{W+\Xi_s(\beta;\Gamma')} \bigg),
\label{Eqn_L_Gamma}
\end{equation}
where $N_0$ is the initial number of spins and $N(\Gamma)$ is the number of spins at scale $\Gamma$. (Note also that we set our initial energy scale to unit, $\Omega_0 = 1$.) At high energy or short distances, $\Pi_s(\beta;\Gamma)$ is close to its UV (ultraviolet) value and we have the random-singlet scaling $\Gamma \sim L^{\psi_0}$ with $\psi_0=\frac{\Pi^0_s}{1+\Pi^0_s}=\frac{1}{3}$, while in the IR (low energy), the scaling between energy and length is modified to $\Gamma \sim L^{\psi_\infty}$ with a different, smaller exponent $\psi_\infty=\frac{\Pi^\infty_s}{1+\Pi^\infty_s}$. As the crossover takes place when $\Omega \sim 1/\beta$, two length scales $L_0 = L_0(\beta)$ and $L_\infty = L_\infty(\beta)$ naturally emerge below and above which the physics is captured by the zero and infinite temperature infinite-randomness fixed points, respectively. This finite-temperature crossover is summarized in Fig.~\ref{Fig_crossover} which shows that at any temperature $T>0$, the system eventually flows to the  infinite-temperature fixed point responsible for the large-scale and long-time behavior, after crossing over from the zero-temperature random-singlet phase.


\begin{figure}[t!]
\includegraphics[width=1.0\columnwidth]{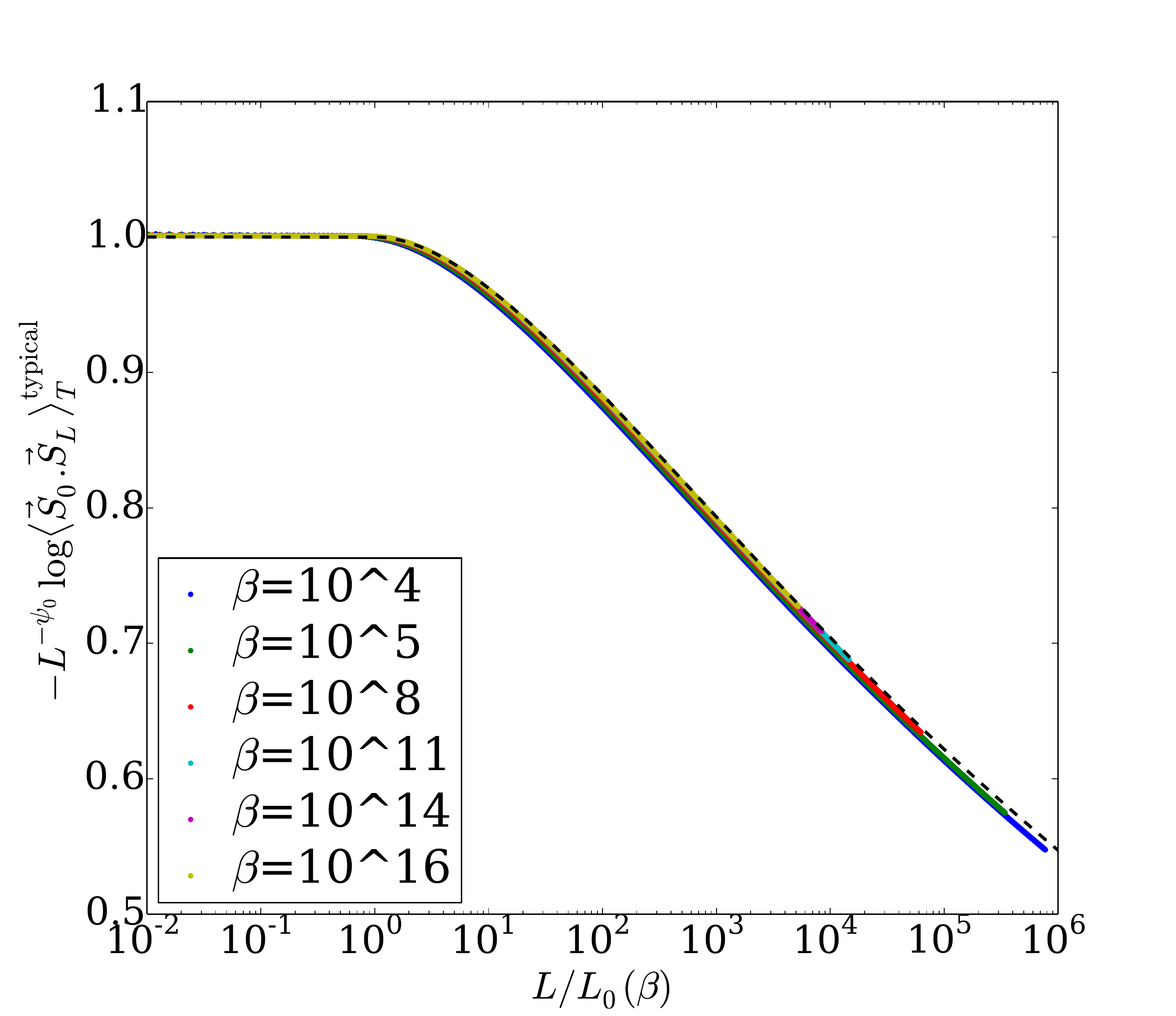}
\caption{Universal scaling function $f(x)$ given by Eq.~\eqref{eqUnivPsi} that controls the typical decay of correlation functions. Data points are obtained by evaluating~\eqref{Eqn_L_Gamma},~\eqref{eqCorrTyp} numerically for various temperatures. The dashed black curve corresponds to Eq.~\eqref{eqExact}.   }
\label{Fig_psi}
\end{figure}

To illustrate how physical quantities scale along the crossover, let us consider for example the typical behavior of the two-point correlation function obtained by measuring the projection operator onto the singlet channel of two sites separated by distance $L$, which we denote by suggestive short-hand notation $\langle \vec{S}_0 \cdot \vec{S}_L \rangle_T$ (with this shorthand notation, the Hamiltonian of our Fibonacci chain can be written as $H =  \sum J_i  \vec{S}_i \cdot \vec{S}_{i+1}$ as for an ordinary Heisenberg spin chain).
 At zero and infinite temperature, the typical behavior of this correlation function is governed by a stretched exponential decay $\sim {\rm e}^{- \alpha L^{\psi_{0,\infty}}}$ (averaged correlation functions should instead decay algebraically because of rare events that dominate the average). At finite temperature, it is given by the typical virtual coupling between sites at distance $L$,
\begin{equation}
\langle \vec{S}_0 \cdot \vec{S}_L \rangle_T^{\rm typical}  \sim J_{\rm typ}(L) = {\rm e}^{- \Gamma(L) -\overline{\xi}(L)},
\label{eqCorrTyp}
\end{equation}
where $\overline{\xi}=W + \Xi_s(\beta;\Gamma)$ and $\Gamma(L)$ is obtained by inverting Eq.~\eqref{Eqn_L_Gamma}. Taking the scaling limit $T \to 0$ and $L\to \infty$ of Eq.~\eqref{eqCorrTyp} numerically while keeping $L/L_0(\beta)$ fixed [recall that $L_0(\beta) \sim (\log \beta)^{1/\psi_0}$], we find that the typical spin-spin correlation function obeys the following universal scaling 
\begin{equation}
\langle \vec{S}_0 \cdot \vec{S}_L \rangle_T^{\rm typical} \sim  \exp \left[ -C L^{\psi_0} f\left(L/L_0(\beta) \right) \right], 
\label{eqUnivPsi}
\end{equation}
where $C$ is a non-universal (cutoff dependent) constant, $\psi_0 = \frac{1}{3}$, and $f(x)$ is a universal scaling function that interpolates between $f(0)=1$ in the UV and $\lim_{x\rightarrow \infty}f(x)\sim x^{\psi_\infty-\psi_0}$ in the IR (Fig.~\ref{Fig_psi}). 

The full functional form of the crossover function $f(x)$ can be computed in closed form by approximating the double-exponential Boltzmann factors of the form $e^{-\beta \Omega_0 e^{-\Gamma}}$ in $\Pi_s(\beta;\Gamma)$ by step functions, from which we find
%
\begin{equation}
f(x) =  
\begin{cases}
1, &  \text{ if } x<1,\\
x^{-\psi_0} \frac{\Pi_s^\infty-\Pi_s^0}{\Pi_s^\infty (1+\Pi_s^0)}+x^{\psi_\infty-\psi_0} \frac{\Pi_s^0(1+\Pi_s^\infty)}{\Pi_s^\infty (1+\Pi_s^0)} , &\text{ if } x>1.
\end{cases}
\label{eqExact}
\end{equation}
This function is plotted in dashed lines in Fig.~\ref{Fig_psi}, and produces close agreement with the numerical evaluation of Eq.~\eqref{Eqn_L_Gamma} using the exact form of $\Pi_s(\beta;\Gamma)$. 

This computation demonstrates the methodology for computing universal crossover scaling of other quantities, including mean correlation functions as well as more complicated observables such as entanglement entropy~\cite{RefaelMoore,FidkowskiPRB08}.

\section{Conclusion \label{secConclusion}}
%
%
In this paper, we studied analytically the universal crossover between ground-state and excited-state quantum criticality in a strongly disordered nonergodic anyonic chain. We argued that the finite ``temperature'' (or energy density) properties of this system can be efficiently captured using RSRG-X using a ``local-node sampling'' that is suitable for analytic treatments. This allowed us to compute exactly universal scaling functions describing eigenstate correlation functions along the crossover, and we expect that other quantities (such as the entanglement entropy) could be computed similarly. Although we focused our analysis on the Fibonacci chain, we expect our main conclusions to hold for more general $SU(2)_k$ anyonic chains with $k \geq 4$. It would be very interesting to find an example of a disordered quantum spin chain with excited-state critical properties analogous to the anyonic chain studied here; we expect that this will require some amount of fine-tuning (recall that at zero temperature, anyonic chains usually allow one to access multicritical points without fine-tuning, for both clean and random couplings). We leave these open questions for future work.   

\smallskip

{\it Acknowledgments.}
We thank S. Parameswaran for many insightful discussions and for collaborations on related matters. This work was supported by NSF Grant No. DMR-1507141, an Investigator Grant from the Simons Foundation, the Korea Foundation for Advanced Studies (B.K.), the Gordon and Betty Moore Foundation's EPiQS Initiative through Grant No. GBMF4307 (A.C.P.), and the LDRD Program of LBNL (R.V.).
\appendix

\section{Disordered Fibonacci anyon chain and RSRG-X}
\label{Appendix_A}

\begin{figure}[h!]
\includegraphics[width=0.5\columnwidth]{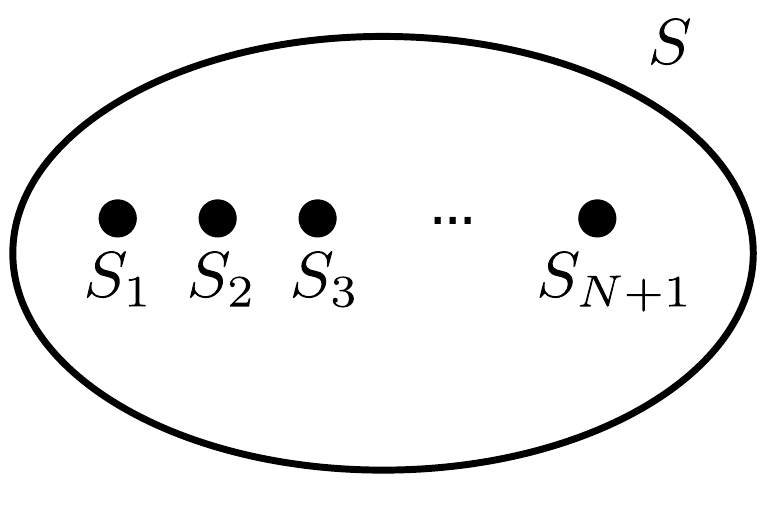}
\caption{A system with $N+1$ anyons on a 2D surface with total topological charge $S$.}
\label{Fig_many_anyons}
\end{figure}

\begin{figure}[h!]
\includegraphics[width=1.0\columnwidth]{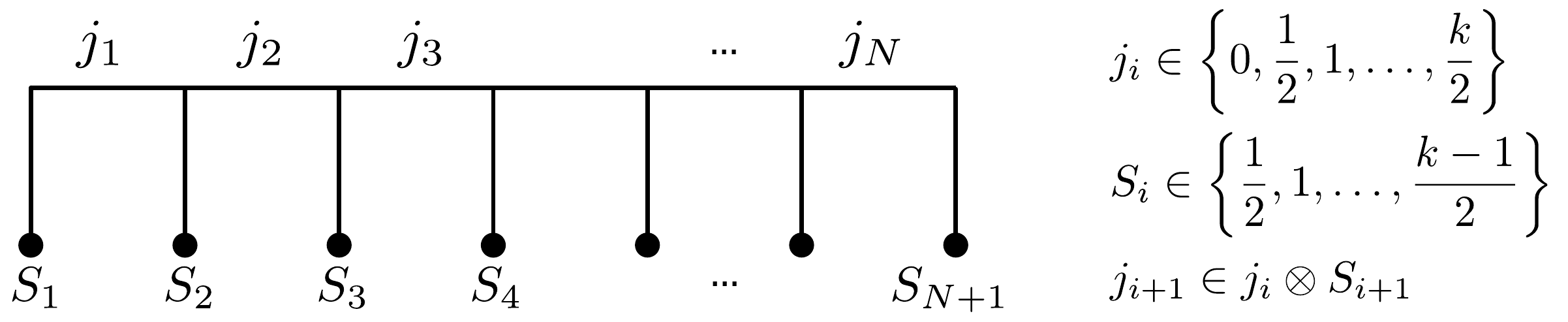}
\caption{Basis vector of the Fibonacci anyon Hilbert space of $N+1$ anyons $S_1,\dots,S_{N+1}$, labeling ``legs'' of the basis. Dots are drawn to provide a connection with dots in Fig.~\ref{Fig_many_anyons}. Each vertex satisfies the admissibility condition (fusion rule).}
\label{Fig_basis}
\end{figure}

\begin{figure}[h!]
\includegraphics[width=1.0\columnwidth]{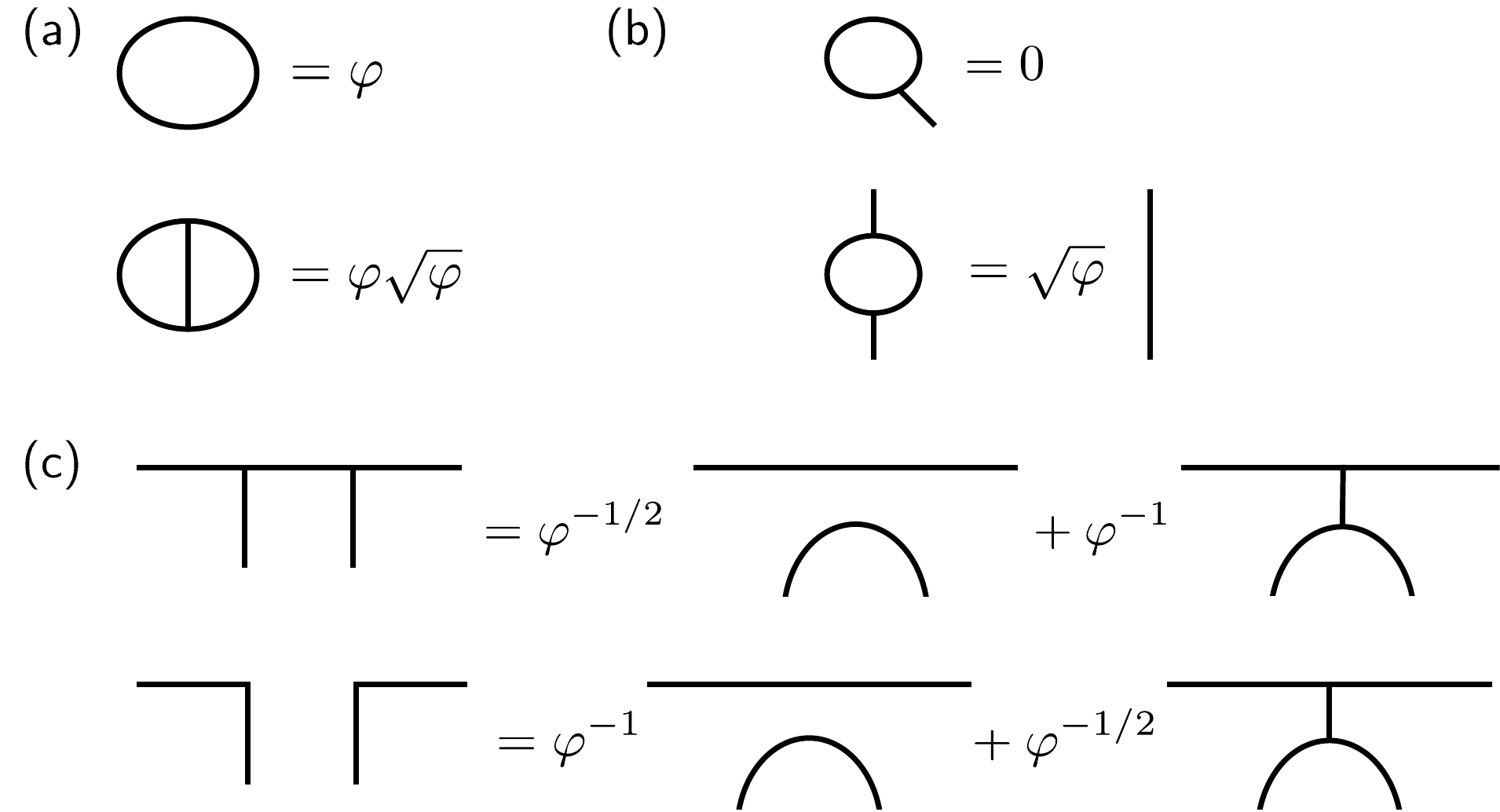}
\caption{(a) Quantum dimension of the Fibonacci anyon system. (b) Useful identities including the `no tadpole' rule. (c) Nontrivial ``F moves'' in the Fibonacci anyon system.}
\label{Fig_Fibonacci_rules}
\end{figure}

\begin{figure}[h!]
\includegraphics[width=0.8\columnwidth]{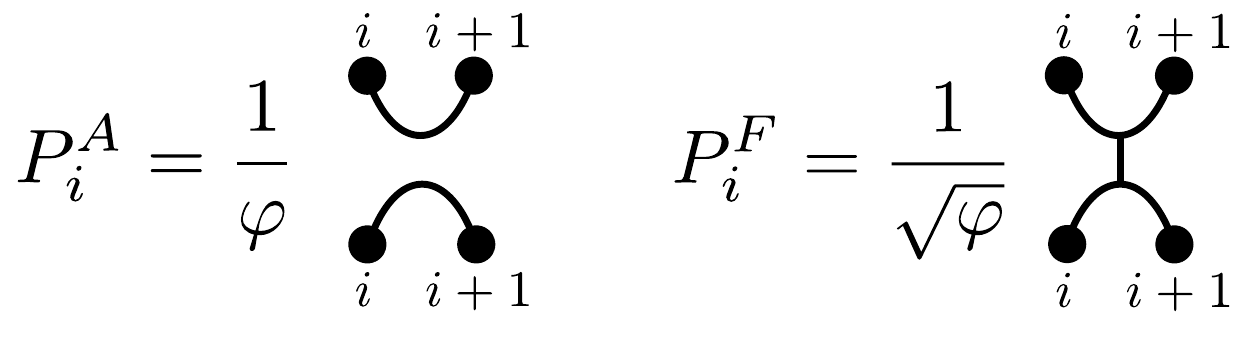}
\caption{Projection operators. Left: Definition of $P_i^A$ which projects spins at $i$ and $i+1$ onto the singlet (trivial anyon). Right: Definition of $P_i^F$ which projects spins at $i$ and $i+1$ onto the triplet ($\tau$ anyon).}
\label{Fig_Fibonacci_projectors}
\end{figure}

\begin{figure}[h!]
\includegraphics[width=1.0\columnwidth]{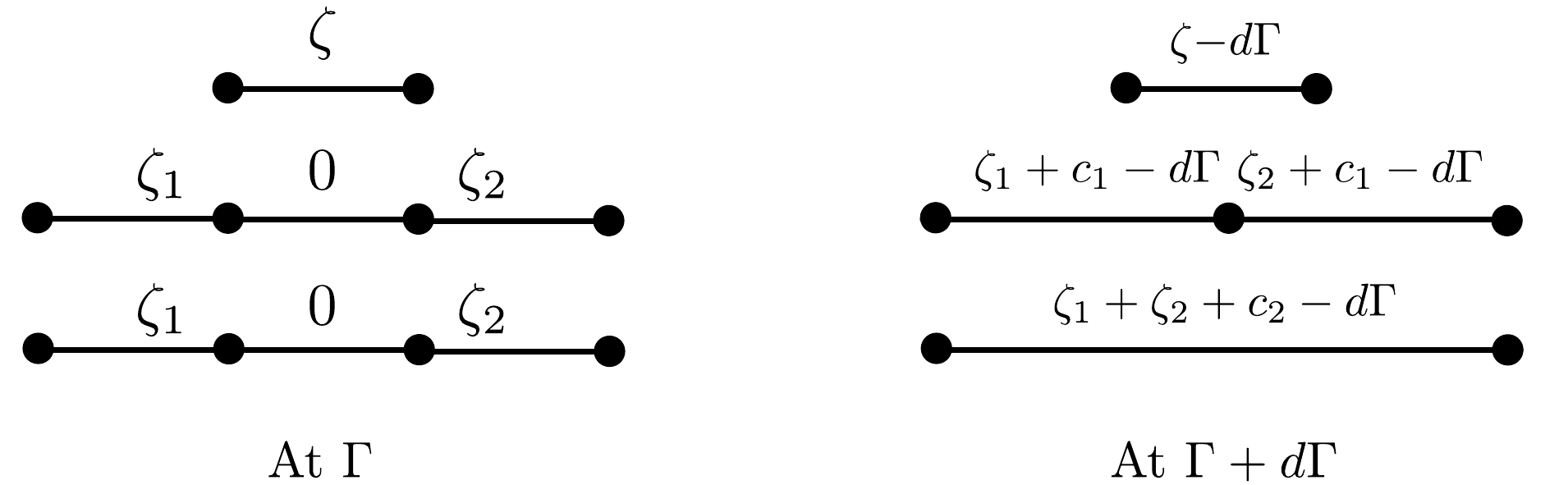}
\caption{Change in logarithmic bond strength upon changing the RG scale from $\Gamma$ to $\Gamma + d\Gamma$. $0$ denotes the strongest bond at $\Gamma$ which is decimated away at $\Gamma+d\Gamma$. Left: Bond strengths at $\Gamma$. Right: Bond strengths at $\Gamma+d\Gamma$. Due to the change in the energy scale, logarithmic-variables are shifted by $d\Gamma$. Diagrams at middle and bottom show how the nearby bonds change as a result of a triplet (ferromagnetic) and a singlet (antiferromagnetic) decimation.}
\label{Fig_renormalized_couplings}
\end{figure}

Formally, the algebraic theory of anyonic systems can be described using the language of modular tensor category~\cite{RevModPhys.80.1083,wang2010topological}. In this appendix, we mainly focus on the (disordered) Fibonacci anyon chain~\cite{wang2010topological,Trebst01062008,PhysRevLett.98.160409}, which is the simplest non-Abelian anyon theory among $SU(2)_{k \ge 2}$ anyon theories. The $SU(2)_k$ anyon theory contains a finite number of particles with labels $\{0,\frac{1}{2},1,\dots,\frac{k}{2} \}$ which are often referred to as ``spins'' of particles. These particles can fuse with each other according to the fusion rules $j_1 \otimes j_2 = |j_1-j_2| \oplus \cdots \oplus \min \big(j_1+j_2, k- (j_1+j_2) \big)$. (In this ``spin'' labeling, $0$ is the trivial or vacuum particle.) The Fibonacci anyon system can be defined as a subcategory of the $SU(2)_3$ anyon theory allowing only for $\{0,1\}$ anyons. In the following, we relabel anyons from $\{0,1\}$ to $\{1,\tau\}$. The fusion rules in the Fibonacci anyon system are $1 \otimes 1 = 1$, $1 \otimes \tau = \tau$, and $\tau \otimes \tau = 1 + \tau$, where the last rule shows the ``non-Abelian'' nature of the Fibonacci anyon.

The Hilbert space of the Fibonacci anyon chain can be constructed as follows. Suppose we place some anyons $S_1,\dots,S_{N+1}$ in a 2D surface where $S_i \in \{ \tau \}$ is the only nontrivial anyon. We can fuse anyons together to yield the \textit{total topological charge} $S$ as described in Fig. ~\ref{Fig_many_anyons}. Because the Fibonacci anyon system is ``non-Abelian'' there could be multiple ways of fusing anyons and this multiplicity becomes the \textit{dimension} of the Hilbert space. To be specific, the Hilbert space is spanned by the (unnormalized) basis $\{ \left| j_1,\dots,j_N \right> : j_i \in \{1,\tau \}, j_{i+1} \in j_i \otimes S_i \}$. We demand that every vertex satisfy the ``admissible condition'' (or fusion rule) $j_{i+1} \in j_i \otimes S_i$. We use a solid line in Fig.~\ref{Fig_basis} to represent the $\tau$ anyon, and an empty line to represent the trivial anyon. From the fusion rules, trivial anyons ($1$'s) cannot appear adjacent to each other: if $j_i = 1$ for some $i$, then we must have $j_{i-1} = j_{i+1} = \tau$. This restriction implies that the dimension of the Hilbert space of $N+1$ anyons is equal to the $N$th Fibonacci number $F_N$, hence the nomenclature. ($F_1=1$, $F_2=1$, and the rest follows from the recursion relation $F_{n+1} = F_{n} + F_{n-1}$.) The dimension of the Hilbert space grows as $\varphi^N$ as $N$ goes to infinity, with $\varphi$ the golden ratio $\frac{1+\sqrt{5}}{2} \approx 1.618$. The ket vector $\left< j_1,\dots,j_N \right|$, the adjoint of $\left| j_1,\dots,j_N \right>$, is the ``upside-down'' diagram of $\left| j_1,\dots,j_N \right>$, and the inner product is given by the diagrammatic concatenation between ``ket'' and ``bra'' diagrams. Two diagrams are considered as identical up to ``isotopy,'' ``F moves,'' evaluation of quantum dimension, etc.~\cite{wang2010topological}. (A partial set of rules for the Fibonacci case is described in Fig.~\ref{Fig_Fibonacci_rules}.)

Having defined the Hilbert space, we can construct local operators acting on the Hilbert space. We consider \textit{local} projection operators $P_i^A$ and $P_i^F$ which are the ``singlet'' and the ``triplet'' projectors acting on the sites $i$ and $i+1$. The definition of the operators in terms of diagram is provided in Fig.~\ref{Fig_Fibonacci_projectors}. 

The Hamiltonian of the Fibonacci anyon chain is given by
\begin{equation}
\label{fibonacci_hamiltonian}
H = -\sum_{i=1}^NJ_i P_i^A 
\end{equation}
where $J_i$ are the interaction strength. When $J_i>0$ ($<0$), the singlet (triplet) fusion is preferred between $S_i$ and $S_{i+1}$ spins to lower the energy.\\

\section{Derivation of the finite-temperature flow equation \label{flow_eqn_derivation}}
In this section, we derive the finite temperature flow equations~\eqref{Finite_T_flow_eqn} using the local node probabilities $(p_A^1,p_A^\tau,p_F^1,p_F^\tau)$ for the different fusion channels. These flow equations describe how the coupling constants change as we decimate the largest energy gaps $ \Omega = \max_{i} |J_i|$  in the system. We introduce the logarithmic variables $\Gamma = \log(\Omega_0/\Omega)$ and $\zeta_i = \log(\Omega/|J_i|)$, where $\Omega_0$ is the initial energy scale. The RG scale $\Gamma$ grows from $0$ to $\infty$ along the flow and $\zeta_i \ge 0$ where the equality holds when $i$ is the strongest bond. Let $\rho_F(\zeta,\Gamma)$ and $\rho_A(\zeta,\Gamma)$ be the probability distribution of having a ferromagnetic ($J_i<0$) and antiferromagnetic bond ($J_i>0$) with bond strength $\zeta$ at scale $\Gamma$. The probability of having \text{(anti-)ferromagnetic} bonds at scale $\Gamma$ is $p_F = \int \rho_F(\zeta,\Gamma) d \zeta$ [$p_A = \int \rho_A(\zeta,\Gamma) d\zeta$], where $p_A+p_F = 1$.

As we increase the RG scale by $d\Gamma$, we decimate $\rho_F(0,\Gamma) d\Gamma$  ferromagnetic bonds and $\rho_A(0,\Gamma) d \Gamma$ antiferromagnetic bonds. Due to this decimation, the coupling of the nearby bonds is renormalized. It is convenient to define the constants $c_1 = \log(\tau)$ and $c_2 = \log(\tau^2/2)$. As the energy scale $\Omega$ changes, the logarithmic coupling strength $\zeta$ also changes. The effects of the change in the coupling constants are summarized in Fig.~\ref{Fig_renormalized_couplings}. With all these ingredients, we can derive the flow equation:
\begin{widetext}
\begin{align}
\rho_F (\zeta, &\Gamma+d \Gamma) d\zeta \nonumber \\
&= \bigg( \rho_F(\zeta + d\Gamma, \Gamma) d \zeta + \rho_F(0,\Gamma) d\Gamma \Big[ 2 p_F^1 \int_0 ^\infty d \zeta_1 \int_0^\infty d \zeta_2 \delta(\zeta - \zeta_1 - \zeta_2 - c_2 +d\Gamma ) \rho_F(\zeta_1 ,\Gamma) \rho_A(\zeta_2, \Gamma) \nonumber \\
&\qquad \qquad \qquad \qquad \qquad \qquad \qquad \qquad \qquad \qquad \qquad \quad +2 p_F^\tau \rho_A(\zeta - c_1 +d\Gamma, \Gamma) d\zeta -2\rho_F(\zeta+d\Gamma, \Gamma) d\zeta \Big] \nonumber \\
&\qquad \qquad \qquad \qquad \quad + \rho_A(0,\Gamma) d\Gamma \Big[ 2p_A^1 \int_0 ^\infty d \zeta_1 \int_0^\infty d \zeta_2 \delta(\zeta - \zeta_1 - \zeta_2 - c_2 +d\Gamma ) \rho_F(\zeta_1 ,\Gamma) \rho_A(\zeta_2, \Gamma) \nonumber \\
&\qquad \qquad \qquad \qquad \qquad \qquad \qquad \qquad \qquad \qquad \qquad \quad + 2 p_A^\tau \rho_A (\zeta-c_1 +d\Gamma, \Gamma) d\zeta -2\rho_F(\zeta+d\Gamma, \Gamma) d\zeta \Big] \bigg) \nonumber \\
&\qquad \times \bigg( 1 - (2p_F^1 + p_F^\tau) \rho_F(0,\Gamma) d\Gamma - (2p_A^1 + p_A^\tau) \rho_A(0,\Gamma) d\Gamma \bigg)^{-1},
\label{flow_eqn_A}
\end{align}
where the first term on the right-hand side is the number of couplings that lie in $\zeta \sim \zeta+ d\zeta$ after the decimation. The second term is the generation of new couplings in $(\zeta, \zeta + d\zeta)$ after decimating ferromagnetic bonds to singlets. When decimating a ferromagnetic bond to a singlet, a ferromagnetic bond is generated if and only if one of the adjacent bond is ferromagnetic and the other adjacent bond is antiferromagnetic. The third term counts the generation of ferromagnetic bonds by antiferromagnetic bonds for which the adjacent ferromagnetic bonds are decimated to triplets, as well as the loss of ferromagnetic bonds due to the decimation. The fourth and the fifth terms count the generation and removal of ferromagnetic bonds due to the decimation of antiferromagnetic bonds. Finally, the last term is the change in the total probability due to the decrease in the total number of couplings. The probability distribution of antiferromagnetic bonds can analogously be obtained:
\begin{align}
\rho_A (\zeta, &\Gamma+d \Gamma) d\zeta \nonumber \\
&= \bigg( \rho_A(\zeta + d\Gamma, \Gamma) d \zeta \nonumber \\
&\qquad  \quad + \rho_F(0,\Gamma) d\Gamma \Big[ p_F^1 \int_0 ^\infty d \zeta_1 \int_0^\infty d \zeta_2 \delta(\zeta - \zeta_1 - \zeta_2 - c_2 +d\Gamma ) \big(\rho_F(\zeta_1 ,\Gamma) \rho_F(\zeta_2, \Gamma) + \rho_A(\zeta_1 ,\Gamma) \rho_A(\zeta_2, \Gamma) \big) \nonumber \\
&\qquad \qquad \qquad \qquad \qquad \qquad \qquad \qquad \qquad \qquad \qquad \qquad \qquad +2 p_F^\tau \rho_F(\zeta - c_1 +d\Gamma, \Gamma) d\zeta -2\rho_A(\zeta+d\Gamma, \Gamma) d\zeta \Big] \nonumber \\
&\qquad \quad + \rho_A(0,\Gamma) d\Gamma \Big[ p_A^1 \int_0 ^\infty d \zeta_1 \int_0^\infty d \zeta_2 \delta(\zeta - \zeta_1 - \zeta_2 - c_2 +d\Gamma ) \big(\rho_F(\zeta_1 ,\Gamma) \rho_F(\zeta_2, \Gamma) + \rho_A(\zeta_1 ,\Gamma) \rho_A(\zeta_2, \Gamma) \big) \nonumber \\
&\qquad \qquad \qquad \qquad \qquad \qquad \qquad \qquad \qquad \qquad \qquad \qquad \qquad + 2 p_A^\tau \rho_F (\zeta-c_1 +d\Gamma, \Gamma) d\zeta -2\rho_A(\zeta+d\Gamma, \Gamma) d\zeta \Big] \bigg) \nonumber \\
&\qquad \times \bigg( 1 - (2p_F^1 + p_F^\tau) \rho_F(0,\Gamma) d\Gamma - (2p_A^1 + p_A^\tau) \rho_A(0,\Gamma) d\Gamma \bigg)^{-1}
\label{flow_eqn_F}
\end{align}
Expanding in $d\zeta$ and $d\Gamma$, and throwing away irrelevant terms, Eq.~(\ref{flow_eqn_A}) and Eq.~(\ref{flow_eqn_F}) reduce to Eq.~(\ref{Finite_T_flow_eqn}).
\end{widetext}

\bibliography{QCG_Finite_T2}
    
\end{document}